\documentclass[aps,prd,preprint, nofootinbib]{revtex4-1}
\usepackage{graphicx}
\usepackage{amsmath,amssymb, graphics, setspace,longtable,array,color}
\usepackage[export]{adjustbox}
\usepackage{commath}
\usepackage[export]{adjustbox}
\usepackage{courier,float}
\usepackage{enumitem}
\usepackage{siunitx}
\usepackage{relsize}
\usepackage{multirow}
\usepackage[acronym]{glossaries}
\usepackage{physics}
\usepackage{todonotes}
\usepackage{rotating}
\usepackage[caption=false,position=top]{subfig}

\usepackage{tikz}
\usetikzlibrary{math,shapes,arrows,calc,intersections,automata,positioning, quotes,decorations}
\usepackage{pgfplots}
\pgfplotsset{compat=newest}
\usetikzlibrary{plotmarks}
\usetikzlibrary{arrows.meta,bending}
\usepgfplotslibrary{patchplots}
\usepackage{grffile}

\newcommand{\delay}[1]{\mathbf{D}_{#1}}
\newcommand{\adv}[1]{\mathbf{A}_{#1}}

\newcommand{\mathsym}[1]{{}}
\newcommand{\unicode}[1]{{}}

\makeatletter
\def\p@subsection{}
\def\p@subsubsection{}
\def\p@paragraph{}
\def\p@subparagraph{}
\makeatother

\makeatletter
\def\l@section{\@dottedtocline{1}{1em}{2em}}
\def\l@subsection{\@dottedtocline{2}{1.5 em}{2em}}
\def\l@subsubsection{\@dottedtocline{3}{2em}{3em}}
\def\l@paragraph{\@dottedtocline{4}{2.5em}{4em}}
\def\l@subparagraph{\normalfont \@dottedtocline{5}{3.5 em}{4 em}}
\makeatother

\usepackage{fancyhdr}

\newlength\figH
\newlength\figW

\usepackage{cleveref}

\begin{document}

\title{Characterization of Time Delay Interferometry combinations for the LISA instrument noise}
\label{sec:tdi-algebra}

\author{Olaf~Hartwig}\email{contact: olaf.hartwig@obspm.fr}\affiliation{\addres}
\affiliation{LNE-SYRTE, Observatoire de Paris, Université PSL, CNRS, Sorbonne Université,
	61 avenue de l’Observatoire, 75014 Paris, France}
\author{Martina~Muratore}\email{contact: martina.muratore@unitn.it}\affiliation{\addressi}

\def\addres{Max-Planck-Institut f\"ur Gravitationsphysik (Albert-Einstein-Institut), Callinstra\ss e 38, 30167 Hannover, Germany}
\def\addressi{Dipartimento di Fisica, Universita di Trento and Trento Institute for 
Fundamental Physics and Application / INFN, 38123 Povo, Trento, Italy}
\date{\today}

\begin{abstract}
Time delay interferometry (TDI) is a post-processing technique used in the Laser Interferometer Space Antenna (LISA) to reduce laser frequency noise by building an equal-arm interferometer via combining time-shifted raw phase measurements. The set of so-called 2nd generation TDI variables which sufficiently suppress laser frequency noise considering realistic LISA orbital dynamics has recently been expanded by a large number of additional solutions. In this paper, we characterize these new TDI channels by relating them to the well-known 1st generation variables $\alpha$, $\beta$, $\gamma$, and $\zeta$. We compute explicitly how each 2nd generation variable can be approximated as a linear combination of these four 1st generation variables, and show numerically that these approximations are accurate enough to model the noises not suppressed by TDI. We use these results to discuss how the newly found channels might be advantagous to use for the LISA data analysis. In addition, we demonstrate that newly found variants of the variable $\zeta$ significantly out-perform the ones previously known from the literature.

\end{abstract}

\maketitle

\section{Introduction}
Gravitational waves (GWs) are predicted by General Relativity and were first observed in 2016 by the two ground-based interferometers LIGO \cite{PhysRevLett.116.061102}. They can be detected by laser interferometers which recover the Doppler shift that a passing GW causes on the frequency of the laser beam. \\ The Laser Interferometer Space Antenna (LISA), which is the 3rd large mission (L3) of the ESA program Cosmic Vision has the goal of detecting gravitational waves with frequencies in the mHz regime by tracking the relative distance between two free-falling test masses, hosted in distant spacecraft, using laser interferometry  \cite{Audley:2017drz}. \\ The constellation is designed to have three identical spacecraft which are separated by 2.5 million \si{\kilo\meter} in a triangular formation, with six active laser links connecting them.
The orbits are chosen to have a constellation forming a triangle as equilateral as possible. However, due to celestial dynamics, the arms differ from each other by $\pm1\%$ and the satellites have a relative drift of up to \SI{10}{\meter\per\second}. As a consequence, the LISA interferometers will have unequal and time-varying arm-lengths, such that they will be strongly affected by laser frequency noise. Thus, to compensate for this noise we apply a post-processing technique called time-delay interferometry (TDI) \cite{Armstrong_1999}. This post-processing technique combines on ground the raw phase meter data by properly time-shifting them in order to build an equivalent equal arm interferometer, insensitive to laser frequency noise \cite{postprocessed-tdi}. Such laser noise suppressing data combinations were first introduced in \cite{Armstrong_1999} for the simplified case of a static constellation, so-called 'first generation TDI'. Combinations accounting for a rigid rotation of the constellation around its center of mass, named 'modified first generation' or '1.5th generation' combinations, were later found in \cite{first-tdi-15}. Finally, combinations accounting for relative velocities between the spacecraft due to the LISA constellations orbital dynamics, called 'second generation TDI', were first presented in \cite{first-tdi-2,second-tdi-2}.

One approach to find such TDI data combinations cancelling laser noise considering realistic orbits is \textit{geometric} TDI, first introduced in \cite{Vallisneri:2005}. This approach was recently revisited in \cite{Muratore_2020}, where new combinations of signals which fulfil the frequency noise suppression requirements were found.
In particular, \cite{Muratore_2020} reports 174 combinations of 16-links and 12 of 12-links while in \cite{Vallisneri:2005}, only 48 of 16-links and none of 12-links were advertised\footnote{We call a 'link' the one-way Doppler measurement between two of the LISA spacecraft. Note that \cite{Vallisneri:2005} also includes solutions of up to 24 links, which were not investigated in \cite{Muratore_2020}.}. These combinations could be reduced to a subset of 35 of 16-links and 3 of 12-links, if one considers as equal combinations that differ by any permutation of the satellites. \\ Moreover, additional combinations of 14-links that were missing in the previous catalogue are reported in \cite{muratore2021time}, where it is illustrated that the total number of combinations could be reduce to a subset of 28 of 16-links, 3 of 14-links and 3 of 12-links, if one considers as equal combinations that differ only by any permutations of satellites or a time reversal symmetry. Including these symmetries, the set grows to a total of 210 distinct combinations up to 16 links.\\

On the other hand, the space of possible TDI solutions can also be described and constructed \textit{algebraically}, at least under certain assumptions. It turns out that the entire space of 1st generation TDI (three different but constant constellation arms) can be generated out of 4 combinations \cite{PhysRevD.65.102002}, while for 1.5th generation TDI variables, 6 fundamental variables are needed \cite{RajeshNayak:2004jzp,Nayak_2005}. Conversely, the general algebraic problem of second generation TDI is up to date still unsolved \cite{Tinto2020}.\\

Lacking a theorem guaranteeing that all 2nd generation variables can be constructed from a finite set of generators, the question arises if some of the 210 combinations contain redundant information and if we are able to find a minimum set of TDI variables containing all the relevant information sufficient to perform the LISA data analysis. To this end, we remark that for the purpose of estimating the coupling of non-suppressed effects, like gravitational waves and most secondary noise sources\footnote{Secondary noises such as test mass and readout noise are not suppressed in the output of TDI combinations, but only modulated by differences of the large delays applied when constructing the combination.}, it is usually sufficient to study TDI under the assumption of 1st generation TDI. The reason is that while small mismatches and dynamic changes in the armlengths have to be taken into account when we aim to reduce laser frequency noise (by several orders of magnitude) they only cause very small corrections to the expressions for the non-suppressed effects. \\ Although it is often considered sufficient to use '0th generation' TDI (meaning three constant and equal arms) to perform data analysis, recent studies showed that neglecting the percent level static arm-length mismatches can in some cases bring large errors when modelling the instrument response to non-suppressed noises and GW signals \cite{muratore2021time}. On the other hand, considering three unequal arms was sufficient there to accurately predict the result of numerical simulations performed taking the full orbital dynamics into account. \\
In this paper, we do a follow up of \cite{muratore2021time}, studying algebraic relationships between the new TDI combinations presented there under the same assumption of first generation TDI. This allows us to highlight how they are connected to the known first-generation variables and suggest which of them might be used for LISA data analysis. \\
The rest of the article is divided in three sections. In \cref{generators}, we report the core combinations found in \cite{muratore2021time} and we illustrate how we can characterize them by simplifying these combinations under the assumptions of 1st generation TDI. Moreover, we compute explicitly how each of these simplified combinations is related to the four generators of first generation TDI, $\alpha$, $\beta$, $\gamma$ and $\zeta$, and discuss the implications for LISA data processing. \\
We then demonstrate numerically in \cref{Dfirstgeneration} that the decompositions of 2nd generation variables into the 1st generation generators are good approximations for the secondary noises. We run simulations without laser frequency noise, which allows us to compute the first generation variables, whereas we include readout noise and test-mass acceleration noise as secondary noises. These simulations show that the approximations are valid to within 3 to 5 orders of magnitude, depending on the Fourier frequency and TDI variable considered. We also discuss advantages and disadvantages of different sets of second generation variables that we can use to represent the first generation generators $\alpha$, $\beta$, $\gamma$ and $\zeta$.\\
Moreover, we analyse the laser noise suppression capabilities of the second generation variable  $\zeta_1$ given in \cite{second-tdi-2} with respect to the new second generation $\zeta$ variables found in \cite{muratore2021time}. The latter show to suppress laser noise several orders of magnitude more than the former and far below the level of secondary noises. Conversely, the residual laser noise of the previously known $\zeta_1$ variable would present a significant noise contribution to the full LISA noise budget.

Finally, we report our conclusion and future perspective in \cref{sec:conclusion}.

\section{\label{generators} Generators of first generation TDI and application to geometric TDI}

The formalism of geometric TDI \cite{Vallisneri:2005,Muratore_2020}
allows to understand physically the properties of TDI combinations, and more practically, enables a systematic search for 2nd generation TDI combinations.

We report in \cref{tab:core-combinations-in-eta} the list of the 34 core combinations\footnote{These 34 core combinations can be used to generate all 210 variables presented as supplementary material in \cite{muratore2021time} by applying the appropriate symmetries and index permutations \cite{Muratore:2021phd, Hartwig:2021phd}. The core combinations $C^{16}_1,C^{16}_4,C^{16}_5,C^{16}_6,C^{16}_7,C^{16}_8,C^{16}_{21}$ and $C^{16}_{22}$ are sufficient to generate all 48 variables presented in \cite{Vallisneri:2005}.} found in \cite{muratore2021time}. We express them in terms of time shifts applied to the intermediary TDI variables $\eta_{ij}$. Note that most of the new variables use not only time-delays, but also time-advancements, as described in \cref{sec:notations}.

The $\eta_{ij}$ are constructed in post-processing from the raw measurements provided by the spacecraft, and correspond to direct tracking of the distance between the two test-masses in a LISA link \cite{Audley:2017drz}. These virtual measurements contain the difference between the laser frequency fluctuations $\phi_{i}$ and $\phi_{j}$ of the local and received laser beams, respectively, where the received beam enters with a time delay by the light travel time $d_{ij}(\tau)$:
	\begin{equation}
    \eta_{ij}(\tau) = \phi_j(\tau - d_{ij}(\tau)) - \phi_i(\tau) + N_{ij}(\tau).
\label{eq:eta-tcb}
\end{equation}
While the laser noise terms $\phi_j$ and $\phi_i$ will be strongly suppressed by TDI, the $N_{ij}$ term summarises any effects not fully suppressed by TDI. In particular, $N_{ij}$ contains unavoidable secondary noises, such as acceleration noise of the test-masses and noise introduced by the optical metrology system, as well as gravitational waves. Note that this is a simplifed model for the LISA measurements, see, e.g., \cite{Hartwig:2021phd} for a recent more detailed description.
\begin{table}[h]
\centering
\resizebox{\textwidth}{!}{
%\tiny
\renewcommand{\arraystretch}{1.3}
\begin{tabular}{| c | c |}
\hline
Name & Expression \\
\hline
$C^{12}_{1}$ & $ \left(1-\delay{13}\delay{32}\delay{21}\right)\eta_{12}+\left(\delay{12}-\delay{13}\delay{32}\delay{21}\delay{12}\right)\eta_{23}+\left(\delay{12}\delay{23}-\delay{13}\delay{32}\delay{21}\delay{12}\delay{23}\right)\eta_{31}$ \\
& $-\left(1-\delay{12}\delay{23}\delay{31}\right)\eta_{13}-\left(\delay{13}-\delay{12}\delay{23}\delay{31}\delay{13}\right)\eta_{32} -\left(\delay{13}\delay{32}-\delay{12}\delay{23}\delay{31}\delay{13}\delay{32}\right)\eta_{21}$
\\ \hline 
$C^{12}_{2}$ & $ \left(\adv{23}\adv{31}-\delay{21}\adv{13}\adv{31}\right)\eta_{12}+\left(\adv{23}\adv{31}\delay{12}-\delay{21}\adv{13}\adv{31}\delay{12}\right)\eta_{23} + \left(\delay{21}\adv{13}-\delay{21}\adv{13}\adv{31}\delay{12}\delay{23}\right)\eta_{31}$ \\ 
& $- \left(\adv{23}\adv{31}-\delay{21}\adv{13}\adv{31}\right)\eta_{13} - \left(\adv{23}-\adv{23}\adv{31}\delay{12}\delay{23}\right)\eta_{32}-\left(1-\adv{23}\adv{31}\delay{12}\delay{23}\delay{32}\right)\eta_{21}$\\ \hline 
$C^{12}_{3}$ & $\left(\delay{32}\delay{23}\adv{31}-\delay{31}\adv{12}\delay{23}\adv{31}\right)\eta_{13} - \left(\delay{31}\adv{12}-\delay{31}\adv{12}\delay{23}\adv{31}\delay{12}\right)\eta_{21}+\left(\delay{31}\adv{12}\delay{23}\adv{31}-\delay{32}\delay{23}\adv{31}\right)\eta_{12} $ \\ 
& $- \left(\delay{32}-\delay{31}\adv{12}\right)\eta_{23}+ \left(1-\delay{32}\delay{23}\adv{31}\delay{12}\adv{23}\right)\eta_{31}- \left(1-\delay{32}\delay{23}\adv{31}\delay{12}\adv{23}\right)\eta_{32}$\\ \hline 
\hline
$C^{14}_{1}$ & $ \left(\adv{23}\adv{31}+\adv{23}\adv{31}\delay{12}\delay{23}\delay{31}-\delay{21}-\delay{21}\delay{12}\adv{23}\adv{31}\right)\eta_{12}-\left(\adv{23}\adv{31}-\delay{21}\delay{12}\adv{23}\adv{31}\right)\eta_{13}-\left(1-\adv{23}\adv{31}\delay{12}\delay{23}\delay{31}\delay{12}\right)\eta_{21} $ \\ 
& $+ \left(\adv{23}\adv{31}\delay{12}-\delay{21}\delay{12}\adv{23}\adv{31}\delay{12}\right)\eta_{23}+ \left(\adv{23}\adv{31}\delay{12}\delay{23}-\delay{21}\delay{12}\adv{23}\adv{31}\delay{12}\delay{23}\right)\eta_{31}- \left(\adv{23}-\delay{21}\delay{12}\adv{23}\right)\eta_{32}$\\ \hline 
$C^{14}_{2}$ & $ \left(\delay{23}\adv{31}\adv{12}\adv{23}\delay{31}-\delay{21}\right)\eta_{12}-\left(\delay{23}\adv{31}-\delay{21}\delay{12}\delay{23}\adv{31}\right)\eta_{13} - \left(\delay{23}\adv{31}\adv{12}\adv{23}-\delay{21}\delay{12}\delay{23}\adv{31}\adv{12}\adv{23}\right)\eta_{32}$ \\ 
& $- \left(1-\delay{21}\delay{12}\delay{23}\adv{31}\adv{12}+\delay{23}\adv{31}\adv{12}-\delay{23}\adv{31}\adv{12}\adv{23}\delay{31}\delay{12}\right)\eta_{21}+ \left(\delay{23}\adv{31}\adv{12}\adv{23}-\delay{21}\delay{12}\delay{23}\adv{31}\adv{12}\adv{23}\right)\eta_{31}+ \left(1-\delay{21}\delay{12}\right)\eta_{23}$\\ \hline 
$C^{14}_{3}$ & $ \left(\delay{23}\adv{31}-\delay{21}\delay{12}\delay{23}\adv{31}+\delay{23}\adv{31}\delay{12}\adv{23}\delay{31}-\delay{21}\right)\eta_{12}-\left(\delay{23}\adv{31}-\delay{21}\delay{12}\delay{23}\adv{31}\right)\eta_{13} - \left(\delay{23}\adv{31}\delay{12}\adv{23}-\delay{21}\delay{12}\delay{23}\adv{31}\delay{12}\adv{23}\right)\eta_{32} $ \\ 
& $- \left(1-\delay{23}\adv{31}\delay{12}\adv{23}\delay{31}\delay{12}\right)\eta_{21}+\left(\delay{23}\adv{31}\delay{12}\adv{23}-\delay{21}\delay{12}\delay{23}\adv{31}\delay{12}\adv{23}\right)\eta_{31} + \left(1-\delay{21}\delay{12}\right)\eta_{23}$\\ \hline 
\hline
$C^{16}_{1}$ & $ \left(1-\delay{13}\delay{31}-\delay{13}\delay{31}\delay{12}\delay{21}+\delay{12}\delay{21}\delay{13}\delay{31}\delay{13}\delay{31}\right)\eta_{12}- \left(1-\delay{12}\delay{21}-\delay{12}\delay{21}\delay{13}\delay{31}+\delay{13}\delay{31}\delay{12}\delay{21}\delay{12}\delay{21}\right)\eta_{13} $ \\ 
& $+ \left(\delay{12}-\delay{13}\delay{31}\delay{12}-\delay{13}\delay{31}\delay{12}\delay{21}\delay{12}+\delay{12}\delay{21}\delay{13}\delay{31}\delay{13}\delay{31}\delay{12}\right)\eta_{21}- \left(\delay{13}-\delay{12}\delay{21}\delay{13}\delay{31}\delay{13}+\delay{13}\delay{31}\delay{12}\delay{21}\delay{12}\delay{21}\delay{13}-\delay{12}\delay{21}\delay{13}\right)\eta_{31}$\\ \hline 
$C^{16}_{2}$ & $ \left(\delay{12}\delay{21}\delay{13}\delay{32}\delay{23}\delay{31}-\delay{13}\delay{32}\delay{21}\delay{12}\delay{21}-\delay{13}\delay{32}\delay{21}+1\right)\eta_{12}+\left(\delay{12}\delay{21}\delay{13}\delay{32}\delay{23}\delay{31}\delay{12}-\delay{13}\delay{32}\delay{21}\delay{12}+\delay{12}-\delay{13}\delay{32}\right)\eta_{21} $ \\ 
& $+\left(\delay{12}\delay{21}\delay{13}\delay{32}\delay{23}-\delay{13}\delay{32}\delay{21}\delay{12}\delay{21}\delay{12}\delay{23}\right)\eta_{31}  + \left(\delay{12}\delay{21}\delay{13}\delay{32}-\delay{13}\delay{32}\delay{21}\delay{12}\delay{21}\delay{12}\right)\eta_{23}- \left(\delay{13}-\delay{12}\delay{21}\delay{13}\right)\eta_{32}- \left(1-\delay{12}\delay{21}\right)\eta_{13}$\\ \hline 
$C^{16}_{3}$ & $\left(1-\delay{13}\delay{32}\delay{21}\delay{12}\delay{21}+\delay{12}\delay{23}\delay{31}-\delay{13}\delay{32}\delay{21}\right)\eta_{12} - \left(\delay{13}\delay{32}-\delay{12}\delay{23}\delay{31}\delay{12}\delay{21}\delay{13}\delay{32}+\delay{13}\delay{32}\delay{21}\delay{12}-\delay{12}\delay{23}\delay{31}\delay{12}\right)\eta_{21}$ \\ 
& $+\left(\delay{12}\delay{23}-\delay{13}\delay{32}\delay{21}\delay{12}\delay{21}\delay{12}\delay{23}\right)\eta_{31} -\left(\delay{13}-\delay{12}\delay{23}\delay{31}\delay{12}\delay{21}\delay{13}\right)\eta_{32} - \left(1-\delay{12}\delay{23}\delay{31}\delay{12}\delay{21}\right)\eta_{13}+\left(\delay{12}-\delay{13}\delay{32}\delay{21}\delay{12}\delay{21}\delay{12}\right)\eta_{23} $\\ \hline 
$C^{16}_{4}$ & $ \left(\adv{12}\adv{21}\adv{13}\adv{31}\delay{12}\delay{21}+\adv{12}\adv{21}\adv{13}\adv{31}-\adv{12}\adv{21}-\adv{13}\adv{31}\adv{13}\adv{31}\right)\eta_{12}+ \left(\adv{13}\adv{31}\adv{13}\adv{31}-\adv{13}\adv{31}\adv{13}\adv{31}\delay{12}\delay{21}-\adv{12}\adv{21}\adv{13}\adv{31}+\adv{13}\adv{31}\right)\eta_{13} $ \\ 
& $- \left(\adv{12}-\adv{12}\adv{21}\adv{13}\adv{31}\delay{12}-\adv{12}\adv{21}\adv{13}\adv{31}\delay{12}\delay{21}\delay{12}+\adv{13}\adv{31}\adv{13}\adv{31}\delay{12}\right)\eta_{21}+ \left(\adv{13}-\adv{13}\adv{31}\adv{13}\adv{31}\delay{12}\delay{21}\delay{13}-\adv{12}\adv{21}\adv{13}+\adv{13}\adv{31}\adv{13}\right)\eta_{31}$\\ \hline 
$C^{16}_{5}$ & $\left(\adv{13}\adv{31}-\delay{12}\delay{21}\adv{13}\adv{31}+\adv{13}\adv{31}\delay{12}\delay{21}\delay{13}\delay{31}-1\right)\eta_{12} -\left(\adv{13}\adv{31}-\adv{13}\adv{31}\delay{12}\delay{21}-\delay{12}\delay{21}\adv{13}\adv{31}+\delay{12}\delay{21}\adv{13}\adv{31}\delay{12}\delay{21}\right)\eta_{13}$ \\ 
& $ - \left(\delay{12}\delay{21}\adv{13}\adv{31}\delay{12}-\adv{13}\adv{31}\delay{12}\delay{21}\delay{13}\delay{31}\delay{12}-\adv{13}\adv{31}\delay{12}+\delay{12}\right)\eta_{21}-\left(\adv{13}-\adv{13}\adv{31}\delay{12}\delay{21}\delay{13}+\delay{12}\delay{21}\adv{13}\adv{31}\delay{12}\delay{21}\delay{13}-\delay{12}\delay{21}\adv{13}\right)\eta_{31} $\\ \hline 
$C^{16}_{6}$ & $\left(\adv{13}\adv{32}\delay{21}\delay{12}\delay{23}\delay{31}-\delay{12}\delay{21}\adv{13}\adv{32}\delay{21}+\adv{13}\adv{32}\delay{21}-1\right)\eta_{12} - \left(\delay{12}-\adv{13}\adv{32}\delay{21}\delay{12}\delay{23}\delay{31}\delay{12}+\delay{12}\delay{21}\adv{13}\adv{32}-\adv{13}\adv{32}\right)\eta_{21} $ \\ 
& $- \left(\adv{13}-\adv{13}\adv{32}\delay{21}\delay{12}\delay{23}+\delay{12}\delay{21}\adv{13}\adv{32}\delay{21}\delay{12}\delay{23}-\delay{12}\delay{21}\adv{13}\right)\eta_{31}-\left(\adv{13}\adv{32}-\adv{13}\adv{32}\delay{21}\delay{12}-\delay{12}\delay{21}\adv{13}\adv{32}+\delay{12}\delay{21}\adv{13}\adv{32}\delay{21}\delay{12}\right)\eta_{23} $\\ \hline 
$C^{16}_{7}$ & $\left(\adv{23}\adv{31}\adv{13}\delay{32}\delay{21}-\delay{21}\adv{13}\adv{31}\adv{13}\delay{32}\delay{21}+\delay{21}\adv{13}\adv{31}-\adv{23}\adv{31}\right)\eta_{13} - \left(1-\adv{23}\adv{31}\adv{13}\delay{32}\delay{21}\delay{13}\delay{32}+\delay{21}\adv{13}\adv{31}\adv{13}\delay{32}-\adv{23}\adv{31}\adv{13}\delay{32}\right)\eta_{21} $ \\ 
& $+ \left(\delay{21}\adv{13}-\delay{21}\adv{13}\adv{31}\adv{13}\delay{32}\delay{21}\delay{13}+\delay{21}\adv{13}\adv{31}\adv{13}-\adv{23}\adv{31}\adv{13}\right)\eta_{31}- \left(\adv{23}-\adv{23}\adv{31}\adv{13}\delay{32}\delay{21}\delay{13}+\delay{21}\adv{13}\adv{31}\adv{13}-\adv{23}\adv{31}\adv{13}\right)\eta_{32}$\\ \hline 
$C^{16}_{8}$ & $\left(\adv{13}\delay{32}\delay{21}\delay{13}\delay{31}+\adv{13}\delay{32}\delay{21}-\delay{13}\delay{32}\delay{21}\adv{13}\delay{32}\delay{21}-1\right)\eta_{13} +\left(\adv{13}\delay{32}\delay{21}\delay{13}\delay{31}\delay{13}\delay{32}-\delay{13}\delay{32}\delay{21}\adv{13}\delay{32}+\adv{13}\delay{32}-\delay{13}\delay{32}\right)\eta_{21} $ \\ 
& $ -\left(\adv{13}-\adv{13}\delay{32}\delay{21}\delay{13}-\delay{13}\delay{32}\delay{21}\adv{13}+\delay{13}\delay{32}\delay{21}\adv{13}\delay{32}\delay{21}\delay{13}\right)\eta_{31} + \left(\adv{13}\delay{32}\delay{21}\delay{13}\delay{31}\delay{13}-\delay{13}\delay{32}\delay{21}\adv{13}+\adv{13}-\delay{13}\right)\eta_{32}$\\ \hline 
$C^{16}_{9}$ & $ \left(\adv{21}\adv{12}\adv{23}\adv{31}\delay{12}\delay{21}+\adv{21}\adv{12}\adv{23}\adv{31}-\adv{23}\adv{31}\adv{13}\adv{32}\delay{21}-\adv{21}\right)\eta_{12}-\left(\adv{21}\adv{12}-\adv{21}\adv{12}\adv{23}\adv{31}\delay{12}-\adv{21}\adv{12}\adv{23}\adv{31}\delay{12}\delay{21}\delay{12}+\adv{23}\adv{31}\adv{13}\adv{32}\right)\eta_{21} $ \\ 
& $+ \left(\adv{23}\adv{31}\adv{13}\adv{32}-\adv{23}\adv{31}\adv{13}\adv{32}\delay{21}\delay{12}\right)\eta_{23} + \left(\adv{23}\adv{31}\adv{13}-\adv{23}\adv{31}\adv{13}\adv{32}\delay{21}\delay{12}\delay{23}\right)\eta_{31}+ \left(\adv{23}\adv{31}-\adv{21}\adv{12}\adv{23}\adv{31}\right)\eta_{13}+ \left(\adv{23}-\adv{21}\adv{12}\adv{23}\right)\eta_{32}$\\ \hline 
$C^{16}_{10}$ & $\left(\adv{23}\adv{31}\delay{12}\delay{21}\delay{13}\delay{31}-\delay{21}\delay{12}\adv{23}\adv{32}\delay{21}-\delay{21}+\adv{23}\adv{31}\right)\eta_{12} - \left(1-\adv{23}\adv{31}\delay{12}\delay{21}\delay{13}\delay{31}\delay{12}+\delay{21}\delay{12}\adv{23}\adv{32}-\adv{23}\adv{31}\delay{12}\right)\eta_{21} $ \\ 
& $+ \left(\adv{23}\adv{31}\delay{12}\delay{21}\delay{13}-\delay{21}\delay{12}\adv{23}\adv{32}\delay{21}\delay{12}\delay{23}\right)\eta_{31} -\left(\adv{23}\adv{31}-\adv{23}\adv{31}\delay{12}\delay{21}\right)\eta_{13} + \left(\delay{21}\delay{12}\adv{23}\adv{32}-\delay{21}\delay{12}\adv{23}\adv{32}\delay{21}\delay{12}\right)\eta_{23}- \left(\adv{23}-\delay{21}\delay{12}\adv{23}\right)\eta_{32}$\\ \hline 
$C^{16}_{11}$ & $\left(\adv{31}-\delay{32}\delay{21}\adv{13}\delay{32}\delay{21}\right)\eta_{12}- \left(\delay{32}-\adv{31}\delay{12}\delay{23}\delay{32}\delay{21}\delay{13}\delay{32}-\adv{31}\delay{12}\delay{23}\delay{32}+\delay{32}\delay{21}\adv{13}\delay{32}\right)\eta_{21}+ \left(\adv{31}\delay{12}-\delay{32}\delay{21}\adv{13}\delay{32}\delay{21}\delay{12}\right)\eta_{23} $ \\ 
& $- \left(\adv{31}-\adv{31}\delay{12}\delay{23}\delay{32}\delay{21}\right)\eta_{13}+\left(\delay{32}\delay{21}\adv{13}-\delay{32}\delay{21}\adv{13}\delay{32}\delay{21}\delay{12}\delay{23}\right)\eta_{31} - \left(1-\adv{31}\delay{12}\delay{23}\delay{32}\delay{21}\delay{13}-\adv{31}\delay{12}\delay{23}+\delay{32}\delay{21}\adv{13}\right)\eta_{32} $\\ \hline 
$C^{16}_{12}$ & $\left(\adv{13}\adv{31} - 1\right)\eta_{12}-\left(\delay{12}\delay{23}\adv{31}\adv{13}\delay{32}\delay{21}-\adv{13}\adv{31}\delay{12}\delay{23}\delay{31}-\delay{12}\delay{23}\adv{31}+\adv{13}\adv{31}\right)\eta_{13} - \left(\delay{12}\delay{23}\adv{31}\adv{13}\delay{32}-\adv{13}\adv{31}\delay{12}\delay{23}\delay{31}\delay{13}\delay{32}\right)\eta_{21} $ \\ 
& $-\left(\delay{12}\delay{23}\adv{31}\adv{13}\delay{32}\delay{21}\delay{13}-\adv{13}\adv{31}\delay{12}\delay{23}-\delay{12}\delay{23}\adv{31}\adv{13}+\adv{13}\right)\eta_{31} - \left(\delay{12}\delay{23}\adv{31}\adv{13}-\adv{13}\adv{31}\delay{12}\delay{23}\delay{31}\delay{13}\right)\eta_{32}- \left(\delay{12}-\adv{13}\adv{31}\delay{12}\right)\eta_{23}$\\ \hline 
$C^{16}_{13}$ & $ \left(\adv{32}-\adv{32}\delay{21}\delay{12}\adv{23}\adv{31}\delay{12}\right)\eta_{23}- \left(\delay{31}\adv{12}\adv{23}\adv{31}-\adv{32}\delay{21}\delay{12}\adv{23}\adv{31}\right)\eta_{13}- \left(\adv{32}\delay{21}\delay{12}\adv{23}\adv{31}-\delay{31}\adv{12}\adv{23}\adv{31}\delay{12}\delay{21}-\delay{31}\adv{12}\adv{23}\adv{31}+\adv{32}\delay{21}\right)\eta_{12}$ \\ 
& $- \left(\adv{32}-\delay{31}\adv{12}\adv{23}\adv{31}\delay{12}\delay{21}\delay{12}-\delay{31}\adv{12}\adv{23}\adv{31}\delay{12}+\delay{31}\adv{12}\right)\eta_{21} + \left(1-\adv{32}\delay{21}\delay{12}\adv{23}\adv{31}\delay{12}\delay{23}\right)\eta_{31}- \left(\delay{31}\adv{12}\adv{23}-\adv{32}\delay{21}\delay{12}\adv{23}\right)\eta_{32}$\\ \hline 
$C^{16}_{14}$ & $ \left(\delay{23}\adv{31}\delay{12}\delay{21}\delay{13}\delay{31}-\delay{21}\delay{12}\delay{23}\delay{32}\delay{21}-\delay{21}+\delay{23}\adv{31}\right)\eta_{12}-\left(1-\delay{23}\adv{31}\delay{12}\delay{21}\delay{13}\delay{31}\delay{12}+\delay{21}\delay{12}\delay{23}\delay{32}-\delay{23}\adv{31}\delay{12}\right)\eta_{21} $ \\ 
& $ + \left(\delay{23}\adv{31}\delay{12}\delay{21}\delay{13}-\delay{21}\delay{12}\delay{23}\delay{32}\delay{21}\delay{12}\adv{23}\right)\eta_{31} - \left(\delay{23}\adv{31}-\delay{23}\adv{31}\delay{12}\delay{21}\right)\eta_{13}- \left(\delay{21}\delay{12}\delay{23}-\delay{21}\delay{12}\delay{23}\delay{32}\delay{21}\delay{12}\adv{23}\right)\eta_{32}+ \left(1-\delay{21}\delay{12}\right)\eta_{23}$\\ \hline 
$C^{16}_{15}$ & $ \left(\delay{13}\adv{32}\adv{21}\adv{12}\delay{23}\delay{31}-\adv{12}\adv{21}\delay{13}\delay{32}\delay{21}+\adv{12}\adv{21}-\delay{13}\adv{32}\adv{21}\right)\eta_{12}+ \left(\delay{13}\adv{32}\adv{21}\adv{12}\delay{23}\delay{31}\delay{12}-\adv{12}\adv{21}\delay{13}\delay{32}-\delay{13}\adv{32}\adv{21}\adv{12}+\adv{12}\right)\eta_{21}$ \\ 
& $+ \left(\delay{13}\adv{32}\adv{21}\adv{12}\delay{23}-\adv{12}\adv{21}\delay{13}\delay{32}\delay{21}\delay{12}\adv{23}\right)\eta_{31} - \left(\delay{13}\adv{32}-\delay{13}\adv{32}\adv{21}\adv{12}\right)\eta_{23}- \left(\adv{12}\adv{21}\delay{13}-\adv{12}\adv{21}\delay{13}\delay{32}\delay{21}\delay{12}\adv{23}\right)\eta_{32}+ \left(1-\adv{12}\adv{21}\right)\eta_{13}$\\ \hline 
$C^{16}_{16}$ & $ \left(\delay{23}\delay{32}\adv{21}\adv{12}\adv{23}-\delay{21}\delay{12}\delay{23}\adv{31}\adv{12}\adv{21}\delay{13}\right)\eta_{31}+ \left(1-\delay{21}\delay{12}\right)\eta_{23}+ \left(\delay{23}\delay{32}\adv{21}\adv{12}\adv{23}\delay{31}+\delay{21}\delay{12}\delay{23}\adv{31}\adv{12}\adv{21}-\delay{23}\delay{32}\adv{21}-\delay{21}\right)\eta_{12} $ \\ 
& $+ \left(\delay{21}\delay{12}\delay{23}\adv{31}-\delay{21}\delay{12}\delay{23}\adv{31}\adv{12}\adv{21}\right)\eta_{13} -\left(1-\delay{23}\delay{32}\adv{21}\adv{12}\adv{23}\delay{31}\delay{12}-\delay{21}\delay{12}\delay{23}\adv{31}\adv{12}+\delay{23}\delay{32}\adv{21}\adv{12}\right)\eta_{21} + \left(\delay{23}-\delay{23}\delay{32}\adv{21}\adv{12}\adv{23}\right)\eta_{32}$\\ \hline 
$C^{16}_{17}$ & $\left(\adv{12}\adv{23}\adv{31}\adv{12}\delay{23}\delay{31}-\adv{13}\adv{31}\adv{12}\delay{23}\delay{31}\right)\eta_{12}- \left(\adv{12}-\adv{13}\adv{31}\adv{12}-\adv{12}\adv{23}\adv{31}\adv{12}\delay{23}\delay{31}\delay{12}+\adv{12}\adv{23}\adv{31}\adv{12}\right)\eta_{21}- \left(\adv{12}\adv{23}-\adv{13}\adv{31}\adv{12}\delay{23}\delay{31}\delay{12}\adv{23}\right)\eta_{32}  $ \\ 
& $+ \left(\adv{13}-\adv{13}\adv{31}\adv{12}\delay{23}-\adv{13}\adv{31}\adv{12}\delay{23}\delay{31}\delay{12}\adv{23}+\adv{12}\adv{23}\adv{31}\adv{12}\delay{23}\right)\eta_{31}+ \left(\adv{13}\adv{31}-\adv{12}\adv{23}\adv{31}\right)\eta_{13}- \left(\adv{13}\adv{31}\adv{12}-\adv{12}\adv{23}\adv{31}\adv{12}\right)\eta_{23}$\\ \hline 
$C^{16}_{18}$ & $\left(\delay{13}\delay{31}\delay{12}\adv{23}-\delay{12}\delay{23}\delay{31}\adv{12}\adv{23}\right)\eta_{32} + \left(1-\delay{13}\delay{31}\delay{12}\adv{23}\adv{31}+\delay{12}\delay{23}\delay{31}\adv{12}\adv{23}\delay{31}-\delay{13}\delay{31}\right)\eta_{12}-\left(1-\delay{13}\delay{31}\delay{12}\adv{23}\adv{31}\right)\eta_{13}  $ \\ 
& $ + \left(\delay{12}-\delay{13}\delay{31}\delay{12}\adv{23}\adv{31}\delay{12}\right)\eta_{23}- \left(\delay{13}\delay{31}\delay{12}\adv{23}\adv{31}\delay{12}\delay{23}-\delay{12}\delay{23}\delay{31}\adv{12}\adv{23}-\delay{12}\delay{23}+\delay{13}\right)\eta_{31}- \left(\delay{12}\delay{23}\delay{31}\adv{12}-\delay{12}\delay{23}\delay{31}\adv{12}\adv{23}\delay{31}\delay{12}\right)\eta_{21}$\\ \hline 
$C^{16}_{19}$ & $ \left(\adv{23}\adv{31}-\delay{21}\adv{13}\delay{32}\delay{23}\adv{31}\right)\eta_{12}+\left(\adv{23}\adv{31}\delay{12}-\delay{21}\adv{13}\delay{32}\delay{23}\adv{31}\delay{12}+\adv{23}\adv{31}\delay{12}\delay{23}\delay{32}-\delay{21}\adv{13}\delay{32}\right)\eta_{23} + \left(\delay{21}\adv{13}-\delay{21}\adv{13}\delay{32}\delay{23}\adv{31}\delay{12}\delay{23}\right)\eta_{31} $ \\ 
& $ -\left(\adv{23}-\adv{23}\adv{31}\delay{12}\delay{23}\delay{32}\delay{23}-\adv{23}\adv{31}\delay{12}\delay{23}+\delay{21}\adv{13}\right)\eta_{32} - \left(1-\adv{23}\adv{31}\delay{12}\delay{23}\delay{32}\delay{23}\delay{32}\right)\eta_{21}- \left(\adv{23}\adv{31}-\delay{21}\adv{13}\delay{32}\delay{23}\adv{31}\right)\eta_{13}$\\ \hline 
$C^{16}_{20}$ & $\left(\adv{23}\adv{31}-\delay{23}\adv{31}\right)\eta_{12}- \left(\delay{23}\adv{31}\delay{12}\delay{21}\adv{13}\delay{32}-\adv{23}\adv{31}\delay{12}\delay{23}\delay{32}-\adv{23}\adv{31}\delay{12}+1\right)\eta_{23}+\left(\delay{23}\adv{31}\delay{12}\delay{21}\adv{13}-\delay{23}\adv{31}\delay{12}\delay{21}\adv{13}\delay{32}\delay{23}\right)\eta_{31}$ \\ 
& $ - \left(\adv{23}\adv{31}-\delay{23}\adv{31}\right)\eta_{13}- \left(\delay{23}\adv{31}\delay{12}\delay{21}\adv{13}-\adv{23}\adv{31}\delay{12}\delay{23}\delay{32}\delay{23}-\adv{23}\adv{31}\delay{12}\delay{23}+\adv{23}\right)\eta_{32} - \left(\delay{23}\adv{31}\delay{12}-\adv{23}\adv{31}\delay{12}\delay{23}\delay{32}\delay{23}\delay{32}\right)\eta_{21}$\\ \hline 
$C^{16}_{21}$ & $\left(1-\delay{32}\adv{21}\adv{13}\delay{32}\delay{21}\delay{12}\adv{23}-\delay{31}\adv{12}\adv{21}\adv{13}+\delay{32}\adv{21}\adv{13}\right)\eta_{31} -\left(1-\delay{32}\adv{21}\adv{13}\delay{32}\delay{21}\delay{12}\adv{23}-\delay{31}\adv{12}\adv{21}\adv{13}+\delay{32}\adv{21}\adv{13}\right)\eta_{32}  $ \\ 
& $+ \left(\delay{31}\adv{12}\adv{21}\adv{13}\delay{32}\delay{21}-\delay{31}\adv{12}\adv{21}-\delay{32}\adv{21}\adv{13}\delay{32}\delay{21}+\delay{32}\adv{21}\right)\eta_{12}- \left(\delay{31}\adv{12}-\delay{31}\adv{12}\adv{21}\adv{13}\delay{32}-\delay{31}\adv{12}\adv{21}\adv{13}\delay{32}\delay{21}\delay{12}+\delay{32}\adv{21}\adv{13}\delay{32}\right)\eta_{21}$\\ \hline 
$C^{16}_{22}$ & $\left(\adv{13}\delay{32}\adv{21}\adv{12}\adv{23}\delay{31}+\delay{12}\delay{21}\adv{13}\delay{32}\adv{21}-\adv{13}\delay{32}\adv{21}-1\right)\eta_{12} - \left(\delay{12}-\adv{13}\delay{32}\adv{21}\adv{12}\adv{23}\delay{31}\delay{12}+\adv{13}\delay{32}\adv{21}\adv{12}-\delay{12}\delay{21}\adv{13}\delay{32}\adv{21}\adv{12}\right)\eta_{21} $ \\ 
& $- \left(\adv{13}-\adv{13}\delay{32}\adv{21}\adv{12}\adv{23}+\delay{12}\delay{21}\adv{13}\delay{32}\adv{21}\adv{12}\adv{23}-\delay{12}\delay{21}\adv{13}\right)\eta_{31}+ \left(\adv{13}-\adv{13}\delay{32}\adv{21}\adv{12}\adv{23}+\delay{12}\delay{21}\adv{13}\delay{32}\adv{21}\adv{12}\adv{23}-\delay{12}\delay{21}\adv{13}\right)\eta_{32}$\\ \hline 
$C^{16}_{23}$ & $ \left(\adv{21}\adv{12}\delay{23}\adv{31}+\adv{21}\adv{12}\delay{23}\adv{31}\delay{12}\delay{21}-\delay{23}\adv{31}\adv{13}\delay{32}\delay{21}-\adv{21}\right)\eta_{12}- \left(\adv{21}\adv{12}-\adv{21}\adv{12}\delay{23}\adv{31}\delay{12}-\adv{21}\adv{12}\delay{23}\adv{31}\delay{12}\delay{21}\delay{12}+\delay{23}\adv{31}\adv{13}\delay{32}\right)\eta_{21} $ \\ 
& $ + \left(\delay{23}\adv{31}\adv{13}-\delay{23}\adv{31}\adv{13}\delay{32}\delay{21}\delay{12}\adv{23}\right)\eta_{31} - \left(\delay{23}\adv{31}\adv{13}-\delay{23}\adv{31}\adv{13}\delay{32}\delay{21}\delay{12}\adv{23}\right)\eta_{32}+ \left(\delay{23}\adv{31}-\adv{21}\adv{12}\delay{23}\adv{31}\right)\eta_{13}- \left(1-\adv{21}\adv{12}\right)\eta_{23}$\\ \hline 
$C^{16}_{24}$ & $ \left(\adv{13}\adv{32}\adv{21}-\adv{12}\adv{23}\delay{31}\adv{12}\adv{21}\right)\eta_{12}- \left(\adv{13}\adv{32}\adv{21}-\adv{12}\adv{23}\delay{31}\adv{12}\adv{21}\right)\eta_{13}- \left(\adv{12}-\adv{12}\adv{23}\delay{31}\adv{12}\adv{21}\delay{13}\delay{32}-\adv{13}\adv{32}\adv{21}\delay{13}\delay{31}\adv{12}+\adv{12}\adv{23}\delay{31}\adv{12}\right)\eta_{21} $ \\ 
& $+ \left(\adv{13}\adv{32}-\adv{13}\adv{32}\adv{21}\delay{13}\delay{31}\adv{12}\right)\eta_{23}+ \left(\adv{13}-\adv{13}\adv{32}\adv{21}\delay{13}\delay{31}\adv{12}\delay{23}+\adv{12}\adv{23}-\adv{13}\adv{32}\adv{21}\delay{13}\right)\eta_{31}- \left(\adv{12}\adv{23}-\adv{12}\adv{23}\delay{31}\adv{12}\adv{21}\delay{13}\right)\eta_{32}$\\ \hline 
$C^{16}_{25}$ & $\left(\delay{13}\adv{32}\adv{23}\adv{31}-1\right)\eta_{12} - \left(\delay{12}\adv{23}\adv{32}\delay{21}\adv{13}\delay{32}-\delay{13}\adv{32}\adv{23}\adv{31}\delay{12}-\delay{12}\adv{23}\adv{32}+\delay{13}\adv{32}\right)\eta_{23}+ \left(\delay{12}\adv{23}\adv{32}\delay{21}\adv{13}-\delay{12}\adv{23}\adv{32}\delay{21}\adv{13}\delay{32}\delay{23}\right)\eta_{31} $ \\ 
& $-\left(\delay{12}\adv{23}\adv{32}-\delay{13}\adv{32}\adv{23}\adv{31}\delay{12}\delay{23}\delay{32}\right)\eta_{21}+ \left(\delay{12}\adv{23}-\delay{12}\adv{23}\adv{32}\delay{21}\adv{13}+\delay{13}\adv{32}\adv{23}\adv{31}\delay{12}\delay{23}-\delay{13}\adv{32}\adv{23}\right)\eta_{32}+ \left(1-\delay{13}\adv{32}\adv{23}\adv{31}\right)\eta_{13}$\\ \hline 
$C^{16}_{26}$ & $ \left(\delay{31}\adv{12}\delay{23}\adv{31}-\delay{32}\delay{21}\delay{12}\delay{23}\adv{31}+\delay{31}\adv{12}\delay{23}\adv{31}\delay{12}\delay{21}-\delay{32}\delay{21}\right)\eta_{12}- \left(\delay{31}\adv{12}\delay{23}\adv{31}-\delay{32}\delay{21}\delay{12}\delay{23}\adv{31}\right)\eta_{13} +\left(\delay{31}\adv{12}-\delay{32}\delay{21}\delay{12}\right)\eta_{23} $ \\ 
& $-\left(\delay{32}-\delay{31}\adv{12}\delay{23}\adv{31}\delay{12}\delay{21}\delay{12}-\delay{31}\adv{12}\delay{23}\adv{31}\delay{12}+\delay{31}\adv{12}\right)\eta_{21} +\left(1-\delay{32}\delay{21}\delay{12}\delay{23}\adv{31}\delay{12}\adv{23}\right)\eta_{31} - \left(1-\delay{32}\delay{21}\delay{12}\delay{23}\adv{31}\delay{12}\adv{23}\right)\eta_{32}$\\ \hline 
$C^{16}_{27}$ & $\left(1-\delay{13}\delay{31}\delay{12}\delay{23}\adv{31}+\delay{12}\adv{23}\delay{31}\adv{12}\delay{23}\delay{31}-\delay{13}\delay{31}\right)\eta_{12}- \left(\delay{12}\adv{23}-\delay{13}\delay{31}\delay{12}\delay{23}\adv{31}\delay{12}\adv{23}\right)\eta_{32}-\left(1-\delay{13}\delay{31}\delay{12}\delay{23}\adv{31}\right)\eta_{13} $ \\ 
& $- \left(\delay{13}\delay{31}\delay{12}-\delay{12}\adv{23}\delay{31}\adv{12}\right)\eta_{23}- \left(\delay{13}\delay{31}\delay{12}\delay{23}\adv{31}\delay{12}\adv{23}-\delay{12}\adv{23}\delay{31}\adv{12}\delay{23}-\delay{12}\adv{23}+\delay{13}\right)\eta_{31}-\left(\delay{12}\adv{23}\delay{31}\adv{12}-\delay{12}\adv{23}\delay{31}\adv{12}\delay{23}\delay{31}\delay{12}\right)\eta_{21} $\\ \hline 
$C^{16}_{28}$ & $ \left(\adv{12}\delay{23}\adv{31}-\adv{13}\adv{31}\delay{12}\delay{23}\adv{31}+\adv{12}\delay{23}\adv{31}\delay{12}\adv{23}\adv{31}-\adv{13}\adv{31}\right)\eta_{12}- \left(\adv{12}\delay{23}\adv{31}\delay{12}\adv{23}-\adv{13}\adv{31}\delay{12}\delay{23}\adv{31}\delay{12}\adv{23}\right)\eta_{32}+ \left(\adv{12}-\adv{13}\adv{31}\delay{12}\right)\eta_{23}$ \\ 
& $+ \left(\adv{13}\adv{31}\delay{12}\delay{23}\adv{31}-\adv{12}\delay{23}\adv{31}-\adv{12}\delay{23}\adv{31}\delay{12}\adv{23}\adv{31}+\adv{13}\adv{31}\right)\eta_{13}+\left(\adv{13}-\adv{13}\adv{31}\delay{12}\delay{23}\adv{31}\delay{12}\adv{23}\right)\eta_{31} -\left(\adv{12}-\adv{12}\delay{23}\adv{31}\delay{12}\adv{23}\adv{31}\delay{12}\right)\eta_{21} $\\ \hline 
\end{tabular}}
\renewcommand{\arraystretch}{1}
\caption{\label{tab:core-combinations-in-eta} List of the 34 core combinations found in \cite{muratore2021time} expressed in terms of time shifts applied to the intermediary TDI variables $\eta_{ij}$.}
\end{table}

We now want to study how these non-suppressed effects appear in the 34 core TDI combinations, and how many of these combinations we need to extract all information contained in our raw measurements.
To this end, we want to express all variables in terms of a finite set of generators. As we introduced, the algebraic problem of second generation TDI is to date still unsolved \cite{Tinto2020}, such that no set of generators is known for this case. Instead, we study these TDI combinations under the assumptions of first generation TDI since, 
as also proposed in \cite{Tinto2020} and empirically demonstrated in \cite{muratore2021time}, this is sufficient to describe the instrumental noise and GW signal response. It is important to stress that in practice, we cannot use first generation TDI variables in the actual data analysis, since they don’t suppress the laser noise sufficiently. \\ 

To avoid confusion, we will use a different notation for the six non-commutative delay operators of second generation, as used in \cref{tab:core-combinations-in-eta}, and the three commutative delay operators of first generation TDI. For the latter, we take inspiration from the literature, e.g. \cite{PhysRevD.65.102002}, and will denote the operators applying the corresponding delays by  $x$, $y$ and $z$. They act on any time dependent function $f(t)$ via
\begin{equation}
	x f(t) = f(t - x^d)\qcomma y f(t) = f(t - y^d)\qand  z f(t) = f(t - z^d)\qc
\end{equation}
where we compute the delays as
\begin{subequations}
\begin{align}
	x^d &= \text{mean}\qty[\frac{d_{23}(t) + d_{32}(t)}{2}] \label{eq:xd-def}\\
	y^d &= \text{mean}\qty[\frac{d_{31}(t) + d_{13}(t)}{2}] \label{eq:yd-def}\\
	z^d &= \text{mean}\qty[\frac{d_{12}(t) + d_{21}(t)}{2}],\label{eq:zd-def}
\end{align}
\end{subequations}
with $d_{ij}(t)$ as the time series of time varying delays estimated from the orbits for a photon received on spacecraft $i$ and emitted from spacecraft $j$. Since we have three satellites, $i$ and $j$ can take the values 1, 2 or 3, with $i\neq j$. The symbol "$\text{mean[\dots]}$" denotes a time average over the timespan of interest
 (usually a couple of hours), such that $x^d$, $y^d$ and $z^d$ are indeed constants.

The corresponding advancements are denoted by $x^{-1}$, $y^{-1}$ and $z^{-1}$, and act as
\begin{equation}
	x^{-1} f(t) = f(t + x^d)\qcomma y^{-1} f(t) = f(t + y^d)\qand  z^{-1} f(t) = f(t + z^d) .
\end{equation}

We can map all second-generation variables presented in \cref{tab:core-combinations-in-eta} to first generation variables by replacing
\begin{subequations}
\begin{align}
	\delay{12} &= \delay{21} \equiv z\qcomma \delay{23} = \delay{32} \equiv x\qcomma \delay{31} = \delay{13} \equiv y\qc \label{eq:delay-replacement}\\
	\adv{12} &= \adv{21} \equiv z^{-1} \qcomma \adv{23} = \adv{32} \equiv x^{-1}\qcomma \adv{31} = \adv{13} \equiv y^{-1} . \label{eq:adv-replacement}
\end{align}
\end{subequations}
%The stuff below needs a bit of a rewrite, since now I actually have a better result then before! No low-freq. approximation anymore, but instead we show equivalence up to a differential delay.

As known from the literature, one important result for 1st generation TDI is that all TDI variables can be constructed from just four generators \cite{PhysRevD.65.102002}. One possible set of generators are the three Sagnac variables $\alpha$, $\beta$, $\gamma$ together with the fully symmetric Sagnac $\zeta$ \cite{Armstrong_1999}. These are given in our notation as:
\begin{subequations}
\begin{align}
\alpha &= \eta_{12}+z \eta_{23}+zx \eta_{31}-\eta_{13}-y \eta_{32}-yz \eta_{21} \label{eq:operator-alpha}\\
\beta &= \eta_{23}+x \eta_{31}+xy \eta_{12}-\eta_{21}-z \eta_{13}-zy \eta_{32} \label{eq:operator-beta}\\
\gamma &= \eta_{31}+y \eta_{12}+yz \eta_{23}-\eta_{32}-x \eta_{21}-xz \eta_{13} \label{eq:operator-gamma}\\
\zeta &= x\eta_{12}+y \eta_{23}+z \eta_{31}-x\eta_{13}-z \eta_{32}-y \eta_{21}  .\label{eq:operator-zeta}
\end{align}
\end{subequations}
In the interpretation of geometric TDI, $\alpha$, $\beta$ and $\gamma$ are simple two-beam interferometers, while $\zeta$ is a 6 beam interferometer, where each spacecraft emits and receives two beams. \\
Following \cite{PhysRevD.65.102002}, we are able to write each first generation TDI combination as
\begin{equation}
	\text{TDI}_k = A'_k \alpha + B'_k \beta + C'_k \gamma + D'_k \zeta,
	\label{eq:tdi-1-generators}
\end{equation}
where $\text{TDI}_k$ is the k'th combination in \cref{tab:core-combinations-in-eta} after simplifying it using \cref{eq:delay-replacement,eq:adv-replacement}, while $A'_k, B'_k, C'_k, D'_k$ are polynomials of the delay operators $x$, $y$ and $z$. This means that since all combinations given in \cref{tab:core-combinations-in-eta} are just linear combinations of $\alpha$, $\beta$, $\gamma$ and $\zeta$ (under the assumptions of first generation TDI), all information we can extract from any of them is, in principle, already encoded in these four generators.

Note that solutions to \cref{eq:tdi-1-generators} are not necessarily unique, since $\alpha, \beta, \gamma $ and $\zeta$ have a time-delay relationship to each other \cite{Armstrong_1999}:
\begin{equation}
	(1 - x y z)\zeta  = (x - yz)\alpha + (y - x z) \beta + (z - x y) \gamma. \label{eq:zeta-alpha-relationship}
\end{equation}

Before explicitly solving \cref{eq:tdi-1-generators} for all core variables, we further simplify the expressions by applying an overall time shift\footnote{We determine $E_k$ by collecting the common factors in front of each $\eta_{ij}$ in \cref{tab:core-combinations-in-eta} after we used \cref{eq:delay-replacement,eq:adv-replacement} to simplify the combinations from 2nd to 1st generation TDI.} $E_k$ to each expression given in \cref{tab:core-combinations-in-eta}, such that we solve the following equation instead of \cref{eq:tdi-1-generators}:
\begin{equation}
	E_k\ 	\text{TDI}_k = A_k \alpha + B_k \beta + C_k \gamma + D_k \zeta.
	\label{eq:tdi-decomp-timeshift}
\end{equation}
We solve \cref{eq:tdi-decomp-timeshift} for each value of $k$ using the computer software \texttt{Mathematica}. Since each of the six $\eta_{ij}$ is independent, \cref{eq:tdi-decomp-timeshift} can be read as a system of six linear equations, which we first solve for four real coefficients $A_k$, $B_k$, $C_k$ and $D_k$. Note that the actual solutions to \cref{eq:tdi-decomp-timeshift} are only those for which $A_k$, $B_k$, $C_k$ and $D_k$ are also valid polynomials in $x$, $y$ and $z$. In particular, we must not allow solutions containing the inverse of a multi term polynomial such as, e.g., $1/(x - y z)$.

We describe in \cref{app:symmetries} how to retrieve the equivalent expressions for the remaining 174 combinations of the full set of TDI variables  from these core combinations. \\
\subsection{Discussion}
\label{ssec:theory-discussion}
Many of the variables in \cref{tab:all-combs-decomposed} are given as one of the generators $\alpha$, $\beta$, $\gamma$ or $\zeta$ with a single difference of delays in front (e.g., $C_1^{12}\approx (1 - xyz)\alpha$ and $C_{28}^{16}\approx (y^2 - z^2)\zeta$). This implies that there are multiple choices of second generation representatives of the first generation generators given in \cref{eq:operator-alpha} to \cref{eq:operator-zeta}. Indeed, each of these representatives response to GWs and secondary noises can be computed by applying different frequency domain transfer functions to the response of the first generation variables. This means that all second generation version of a generator should have the same signal to noise ratio (SNR). As argued in \cite{Tinto2020}, \cref{eq:zeta-alpha-relationship} further implies that the response of $\zeta$ can be written in terms of that of $\alpha$, $\beta$ and $\gamma$, such that theoretically, one could consider just three variables for the astrophysical data analysis. These could be, for example, second generation versions of $\alpha$, $\beta$ and $\gamma$. Out of these one could construct the quasi-orthogonal channels $A = \frac{\gamma - \alpha}{\sqrt{2}}$, $E = \frac{\alpha - 2 \beta + \gamma}{\sqrt{6}}$ and $T = \frac{\alpha + \beta + \gamma}{\sqrt{3}}$, as first given in \cite{PhysRevD.66.122002}, of which $A$ and $E$ are more sensitive to GWs, while $T$ acts a 'null' channel. We remark that, following \cref{eq:zeta-alpha-relationship}, $T$ is strongly related to $\zeta$, but has a more complicated transfer function. We would therefore suggest that the set $A$, $E$, $\zeta$, which can be easily shown to also be quasi-orthogonal, might be preferable to $A$, $E$, $T$. We discuss the potential impact of these transfer functions in more detail in \cref{Dfirstgeneration}. \\
 Furthermore, since the $\zeta$ combination has the special properties of being less sensitive to gravitational waves compared to $\alpha$, $\beta$ and $\gamma$, at least at low frequencies, we would expect this property to extend to $C^{12}_{3}$, $C^{14}_{3}$, $C^{16}_{26}$, $C^{16}_{27}$ and $C^{16}_{28}$, which are directly related to the first generation $\zeta$. This could make these variables useful for characterising the instrumental noise in the presence of a GW signal, which might be useful to observe stochastic GW backgrounds, as discussed in \cite{muratore2021time}. Note that, as discussed in \cref{Dsecondgeneration} and shown in \cref{laserred}, these combinations  suppress laser noise to the same level as all other variables, contrary to the second generation version of $\zeta$ proposed in \cite{second-tdi-2}. \\
 
Besides the generators $\alpha$, $\beta$, $\gamma$ and $\zeta$, we can also identify second generation versions of other well known first generation variables. For example, the first generation Michelson combination fulfills $X = \qty(\alpha -z \beta -y \gamma +y z  \zeta)$, such that $C_{1}^{16}$, $C_{4}^{16}$ and $C_{5}^{16}$ are possible second generation versions of it, with $C_{1}^{16}$ being the 'standard' second generation version found in the literature. Following \cite{PhysRevD.66.122002}, the Relay variables can be written as $y \alpha - \gamma$, which we identify\footnote{The term $z \alpha - \beta$ appearing in $C^{16}_{6}$ is equal to $y \alpha - \gamma$ under a mirror symmetry, see \cref{app:symmetries}.} in $C^{16}_{6}$, $C^{16}_{7}$, and $C^{16}_{8}$. Lastly, the Monitor variables have the decomposition $\gamma - z \zeta$, such that $C^{16}_{21}$, $C^{16}_{22}$ are second generation versions of them. This is compatible with the classification already presented in \cite{Vallisneri:2005}. However, our table does not contain the Beacon variables identified in \cite{Vallisneri:2005}, since they transform to Monitor variables under a time-reversal symmetry and are therefore omitted in our list of core variables. See \cref{app:symmetries} for more information.
 
 The remaining variables don't seem to have an obvious relationship to other known first generation variables.

In addition, we observe that $C^{16}_{4}$, $C^{16}_{24}$ and $C^{16}_{28}$ contain an overall difference term $(y-z)$ or $(y^2-z^2)$. These terms are vanishing if all delays are assumed equal (i.e., assume $x = y = z$) as explicitly shown for the '0th generation' expressions in \cref{tab:all-combs-decomposed}. This means that when we take the real LISA orbital dynamics into account, the secondary noises as well as the astrophysical signal will be strongly suppressed in these variables. However, they are not exactly vanishing, which highlights that the assumption of equal arms is not sufficient to accurately model the response of some variables.\\

\begin{table}[h]
	\centering
	\small
	\begin{tabular}{| c || c | c || c | c | }
		\hline
		Name & Timeshift & Expression '1st' gen& Timeshift & Expression '0th' gen\\
		\hline
		$C^{12}_{1}$ & $1$ & $(1-x y z) \alpha$ & $1$ & $(1-D^3) \alpha$ \\ \hline 
		$C^{12}_{2}$ & $x y^2$ & $(y-x z) \alpha$ & $ D^2$ & $(1-D) \alpha$ \\ \hline 
		$C^{12}_{3}$ & $y z$ & $(y-x z) \zeta$& $ D$ & $(1-D) \zeta$  \\ \hline 
		\hline
		$C^{14}_{1}$ & $x y$ & $\left(1-z^2\right) \alpha$& $D^2$ & $\left(1-D^2\right) \alpha$ \\ \hline 
		$C^{14}_{2}$ & $y z$ & $\left(1-z^2\right) \gamma$ & $D^2$ & $\left(1-D^2\right) \gamma$ \\ \hline 
		$C^{14}_{3}$ & $y$ & $\left(1-z^2\right) \zeta$ & $D$ & $\left(1-D^2\right) \zeta$  \\ \hline 
		\hline
		$C^{16}_{1}$ & $1$ & $\left(1-y^2 z^2\right) \qty(\alpha -z \beta -y \gamma +y z  \zeta)$ & $1$ & $\left(1-D^4\right) \qty(\alpha -D \beta -D \gamma +D^2  \zeta)$ \\ \hline 
		$C^{16}_{2}$ & $1$ & $\left(1-x y z^3\right) \alpha -z (1-x y z) \beta$ & $1$ & $\left(1-D^5\right) \alpha - (D-D^4) \beta$\\ \hline 
		$C^{16}_{3}$ & $1$ & $\left(1-x y z^3\right) \alpha$ & $1$ & $\left(1-D^5\right) \alpha$ \\ \hline 
		$C^{16}_{4}$ & $y^4 z^2$ & $(y^2-z^2)  \qty( \alpha - z  \beta -y \gamma +y z  \zeta)$ & $D^6 $ & $0$\\ \hline 
		$C^{16}_{5}$ & $y^2$ & $\left(1-z^2\right) \qty(\alpha -z \beta -y \gamma +y z \zeta)$ & $D^2$ & $\left(1-D^2\right) \qty(\alpha -D \beta -D \gamma +D^2 \zeta)$ \\ \hline 
		$C^{16}_{6}$ & $x y$ & $ \left(1-z^2\right) \qty(z\alpha - \beta)$ & $D^2$ & $ \left(1-D^2\right) \qty(D\alpha - \beta)$\\ \hline 
		$C^{16}_{7}$ & $x y^3$ & $(y-x z)\qty(y  \alpha - \gamma)$ & $D^3$ & $(1-D)\qty(D  \alpha - \gamma)$\\ \hline 
		$C^{16}_{8}$ & $y$ & $(1-x y z)\qty(y  \alpha - \gamma)$ & $D$ & $(1-D^3)\qty(D  \alpha - \gamma)$ \\ \hline 
		$C^{16}_{9}$ & $x^2 y^2 z^2$ & $\left(x y-z^3\right) \alpha +\left(z^2-x y z\right) \beta$ & $D^4$ & $\left(1-D\right)( \alpha +\beta)$\\ \hline 
		$C^{16}_{10}$ & $x^2 y$ & $\left(x-y z^3\right) \alpha +\left(y z^2-x z\right) \beta$ & $D^2$ & $\left(1-D^3\right) \alpha +\left(D^2-D\right) \beta$\\ \hline 
		$C^{16}_{11}$ & $y$ & $\left(1-x^2 z^2\right) \alpha$ & $D$ & $\left(1-D^4\right) \alpha$ \\ \hline 
		$C^{16}_{12}$ & $y^2$ & $(1-x y z) \alpha +(x z-y) \gamma$ & $D^2$ & $(1-D^3) \alpha +(D^2-D) \gamma$ \\ \hline 
		$C^{16}_{13}$ & $x^2 y z$ & $\left(x y-z^3\right) \alpha$ & $D^2 $ & $\left(1-D\right) \alpha$ \\ \hline 
		$C^{16}_{14}$ & $y$ & $\left(x y z^2-z\right) \gamma +\left(1-x y z^3\right) \zeta$& $D$ & $\left( D^4-D\right) \gamma +\left(1-D^5\right) \zeta$ \\ \hline 
		$C^{16}_{15}$ & $x z^2$ & $(x y-z) \gamma +(1-x y z) \zeta$& $D^3$ & $(D^2-D) \gamma +(1-D^3) \zeta$ \\ \hline 
		$C^{16}_{16}$ & $y z^2$ & $\left(x y-z^3\right) \gamma +\left(z^2-x y z\right) \zeta$& $D$ & $\left(1-D\right) (\gamma + \zeta)$ \\ \hline 
		$C^{16}_{17}$ & $x y^2 z^2$ & $(y-x z) \beta$& $D^4$ & $(1-D) \beta$ \\ \hline 
		$C^{16}_{18}$ & $x$ & $(x-y z) \alpha$ & $1$ & $(1-D) \alpha$\\ \hline 
		$C^{16}_{19}$ & $x y^2$ & $\left(y-x^3 z\right) \alpha$ & $ D^2$ & $\left(1-D^3\right) \alpha$  \\ \hline 
		$C^{16}_{20}$ & $x y^2$ & $\left(y-x^3 z\right) \alpha +\left(x^2 z-x y\right) \zeta$ & $ D^2$ & $\left(1-D^3 \right) \alpha +\left(D^2 -D \right) \zeta$\\ \hline 
		$C^{16}_{21}$ & $y z^2$ & $(x z-y)\qty( \gamma-z  \zeta)$& $D^2$ & $(D-1)\qty( \gamma-D  \zeta)$ \\ \hline 
		$C^{16}_{22}$ & $y z^2$ & $\left(1-z^2\right)\qty( \gamma -z  \zeta)$ & $D^3$ & $\left(1-D^2\right)\qty( \gamma -D  \zeta)$\\ \hline 
		$C^{16}_{23}$ & $y^2 z^2$ & $\left(x z^2-y z\right) \gamma +\left(y-x z^3\right) \zeta$ & $D^3$ & $\left(D^2-D\right) \gamma +\left(1- D^3\right) \zeta$\\ \hline 
		$C^{16}_{24}$ & $x y z^3$ & $\left(z^2-y^2\right) \alpha$ & $D^5$ & $0 $\\ \hline 
		$C^{16}_{25}$ & $x^2 y$ & $(y-x z) \alpha +(z-x y) \zeta$ & $D^2$ & $(1-D)( \alpha +\zeta)$ \\ \hline 
		$C^{16}_{26}$ & $y z$ & $\left(y-x z^3\right) \zeta$ & $ D$ & $\left(1- D^3\right) \zeta$ \\ \hline 
		$C^{16}_{27}$ & $x$ & $(1-x y z) \zeta$& $D$ & $(1-D^3) \zeta$ \\ \hline 
		$C^{16}_{28}$ & $y^3 z$ & $(y^2-z^2) \zeta$& $D^4$ & $0$ \\ \hline 
	\end{tabular}
	\caption{\footnotesize{\label{tab:all-combs-decomposed} Decomposition of variables from \cref{tab:core-combinations-in-eta} into generators of first generation TDI. Only valid in the approximation of three unequal constant arms. 'Timeshift' denotes the delay to be applied to the combination constructed from the algorithm given in \cite{Muratore_2020}, i.e., the factor $E$ in \cref{eq:tdi-decomp-timeshift}}. We also report how these expression further simply if one assumes all arms to be equal (0th generation TDI), i.e., when assuming $x = y = z \equiv D$. Note that some variables cancel exactly under this assumption.}
\end{table}

\section{\label{verf}Numerical simulations}
We run two simulations using \texttt{LISA Instrument}\footnote{\texttt{LISA Instrument} is a time-domain LISA simulator developed mainly by J.-B. Bayle inside the LISA Consortium. See \cite{simulation-model, Bayle:2019phd, Hartwig:2021phd} for a detailed description of its simulation model.} to verify our statement that the decompositions shown in \cref{tab:all-combs-decomposed} are good approximations for the secondary noises. In both simulations, all noise time series are generated at a high sampling rate of 16 Hz and then filtered and decimated to a lower measurement rate of 4 Hz. The filter is a digital symmetrical FIR filter build from a Kaiser windowing function, with a transition band extending from $\SI{1.1}{\hertz}$ to $\SI{2.9}{\hertz}$ and a minimum attenuation above \SI{2.9}{\hertz} of \SI{320}{\decibel}.\\

In the first simulation, we disable laser noise, which allows us to use the first generation generators $\alpha$, $\beta$, $\gamma$ and $\zeta$. We simulate readout noise and test-mass acceleration noise, as they are the main noise contributors of the instrumental noise after TDI, and use realistic orbits provided by ESA [W. Martens and E. Joffre, personal communication, November 2020] to compute the light travel times. This allows us to directly test the validity of the decomposition given in \cref{tab:all-combs-decomposed} for all the combinations. \\

In the second simulation, we simulate laser noise as well, such that we are restricted to use only the second generation variables. In detail, we demonstrate that the newly identified second generation version of $\zeta$ are significantly more capable of reducing laser frequency noise than those previously known from the literature.\\ 

In \cref{decompositionofTDI}, we use the same simulation to demonstrate numerically that the previously described approximations for the response to unsuppressed effects remain valid also in the presence of laser noise. We show that a set of four 2nd generation combinations can be used to generate the instrument noise response of the channel $C_1^{16}$ (up to an overall transfer function), with a relative error of less than $10^{-2}$ to $10^{-5}$,  depending on the Fourier frequency considered. \\ All TDI combination are computed using \texttt{PyTDI}\footnote{\texttt{PyTDI} is a python package designed to compute TDI combinations, developed mainly by M. Staab and J.-B. Bayle inside the LISA consortium.}.

\subsection{\label{Dfirstgeneration}Simulations without laser noise}
%\subsubsection{Description of simulation. What do we compute?}
We simulate $10^5$ samples of LISA data with the aforementioned parameters, and compute the response of all TDI combinations given in \cref{tab:core-combinations-in-eta}. Here, we use the time-varying light travel times $d_{ij}(t)$ output by the simulation to compute the exact response of the variable given realistic orbits. 

\begin{figure}
	\caption{ Secondary noises in  $C^{12}_1$, $C^{12}_3$, $C^{14}_1$,  $C^{16}_1$ and $C^{16}_{24}$ compared to the residual between the approximation given in \cref{tab:all-combs-decomposed} and their exact expression given in \cref{tab:core-combinations-in-eta}.	\label{fig:algebraic-demo}}
	\begin{center}
		\includegraphics[width=\textwidth]{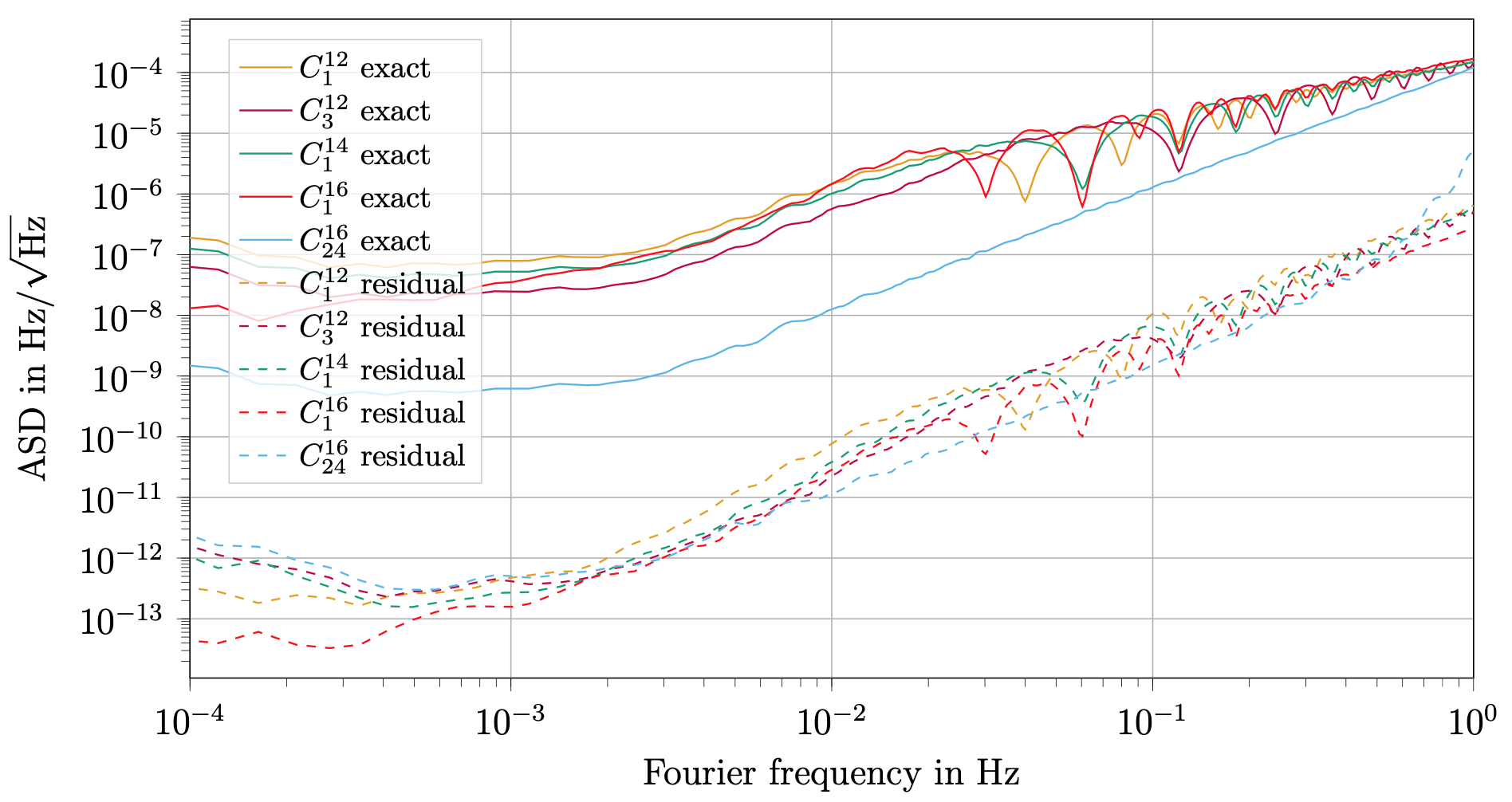}
	\end{center}
\end{figure}

We then compute the constant delays $x^d$, $y^d$ and $z^d$ using \cref{eq:xd-def,eq:yd-def,eq:zd-def}, and use them to construct the generators of first generation TDI as given in \cref{generators}. This allows us to use the expressions given in \cref{tab:all-combs-decomposed} to construct the approximate versions of all TDI combinations. We can then study the time-domain residual between the exact and approximate version of each variable.\\
As an example, we plot in \cref{fig:algebraic-demo} the amplitude spectral density (ASD) of the TDI combinations $C^{12}_1$, $C^{12}_3$, $C^{14}_1$, $C^{16}_1$, and $C^{16}_{24}$ as given in \cref{tab:core-combinations-in-eta}. In addition, we also plot the ASD of the time-domain residual between them and their expressions in terms of $\alpha$, $\beta$, $\gamma$ and $\zeta$ from \cref{tab:all-combs-decomposed}. We observe that the residuals for all variables are at a similar level, about four to five orders of magnitude below the actual secondary noise levels of $C^{12}_1$, $C^{12}_3$, $C^{14}_1$ and $C^{16}_1$, such that the expressions given in \cref{tab:all-combs-decomposed} should provide good approximations of the secondary noises. \\

Inspecting \cref{fig:algebraic-demo}, we observe that the combinations show to have different noise shapes, in particular, their spectra have an unequal number of zeros. This can be explained by referring to the column of \cref{tab:all-combs-decomposed} which shows their approximation in case of equal arms ('0th' generation). We see that all variables contain a difference term of the form $(1 - D^N)$, where $D$ is the single delay operator of '0th' generation TDI, acting as $D f(t) = f(t - \bar d)$, with $\bar d\approx\SI{8.3}{\second}$ as the average arm length. In the frequency domain, this term corresponds to a transfer function of the form $2 \abs{\sin(\pi f N \bar d)}$, with zeros at frequencies at integer multiples of $f =  (N \bar{d})^{-1}$. 

For example, $C^{16}_1$ has $N = 4$, thus shows to have zeros starting at $f \approx \SI{30}{\milli\hertz}$. It is then followed by $C^{12}_1$ with $N=3$,  $C^{14}_1$ with $N=2$ and $C^{12}_3$ with $N=1$. We want to point out that variables with $N = 1$ will have their first zeros starting at roughly $\SI{0.12}{\hertz}$, which is outside the required range of the LISA frequency band, which spans from $\SI{0.1}{\milli\hertz}$ to $\SI{0.1}{\hertz}$ \cite{scird2018}. 

As already argued in the literature (e.g.,  \cite{Vallisneri:2005,Tinto2020}), these additional second generation transfer functions act equally on signal and noise, such that they should not impact the SNR. However, it is argured in \cite{Vallisneri:2005} that they might still negatively impact the data analysis in practice. The reasoning is that the noise level at these zeros will at some point be dominated by other uncanceled noise sources which do not share the same transfer function, such as residual laser frequency noise. We will quantify the fundamental limit of residual laser frequency noise for some exemplary variables in \cref{app:laser-noise-at-zeros}, and show that it only becomes dominant in a negligible frequency interval. However, we cannot exclude that other noise sources with different transfer functions might become dominant in more significant bandwidths close to the zeros. 

In addition, it is pointed out in \cite{Vallisneri:2005} that these zeros can be a challenge for the practical implementation of the data analysis methods, since a large dynamic range is required to properly represent them. It is currently expected that the actual LISA data analysis pipelines can be implemented in such a way that they properly account for these additional in-band zeros, such that there will be no notable impact on the final LISA science products\footnote{S. Babak, personal communication, 2022}. Still, we remark that using combinations such as $C_2^{12}$ and $C_3^{12}$, which do not have any additional zeros inside the LISA requirement band, might be a simpler alternative with respect to specialized treatment of the zeros.\\

The remaining variable shown in \cref{fig:algebraic-demo}, $C^{16}_{24}$, behaves differently from the others, since it is exactly vanishing when assuming equal arms. However, it is not vanishing in the assumption of 1st generation TDI, where it contains a difference term $z^2 - y^2$. This corresponds to a frequency domain transfer function of $2 \abs{\sin(2\pi f (z^d - y^d))}$. With our orbital simulation parameters, the corresponding first zeros are at $f = 0.5 (z^d - y^d)^{-1} \approx \SI{5}{\hertz}$. This is not only above the required LISA band, but also above the stated mission \textit{goals} for the LISA frequency band, which extends to \SI{1}{\hertz} \cite{scird2018}. This transfer function explains why the blue curve describing $C^{16}_{24}$ in \cref{fig:algebraic-demo} appears at a lower level then those of $C^{12}_1$, $C^{12}_3$, $C^{14}_1$ and $C^{16}_1$, since the transfer function starts suppressing the output for frequencies below $\SI{5}{\hertz}$, whereas for the others, the roll-of starts only around \SI{5E-2}{\hertz}.\\

However, we are hestitant to recommend these variables for the actual data analysis since their response strongly depends on the orbits, and can be completely vanishing if two of the armlengths ever share exactly the same value. In addition, the fact that signal and secondary noise are strongly suppressed means that any secondary effects which do not share the $2 \abs{\sin(2\pi f (z^d - y^d))}$ transfer function could become dominant in large parts of the frequency band. In addition, the dynamic range between the input and output data is larger for these variables, such that they might be more susceptible to numerical artifacts introduced during the TDI processing.\\

\begin{figure}
	\caption{Second generation variables as approximation of the derivative of $\alpha$. $C_1^{12}$, $C_2^{12}$, $C_1^{14}$, $C_3^{16}$ and $C_{24}^{16}$ are rescaled by the numerical value of the difference of delays applied in \cref{tab:all-combs-decomposed}. Combinations with larger delay differences have more zeros inside the LISA band.\label{fig:alpha-demo}}	
	\begin{center}
		\includegraphics[width=\textwidth]{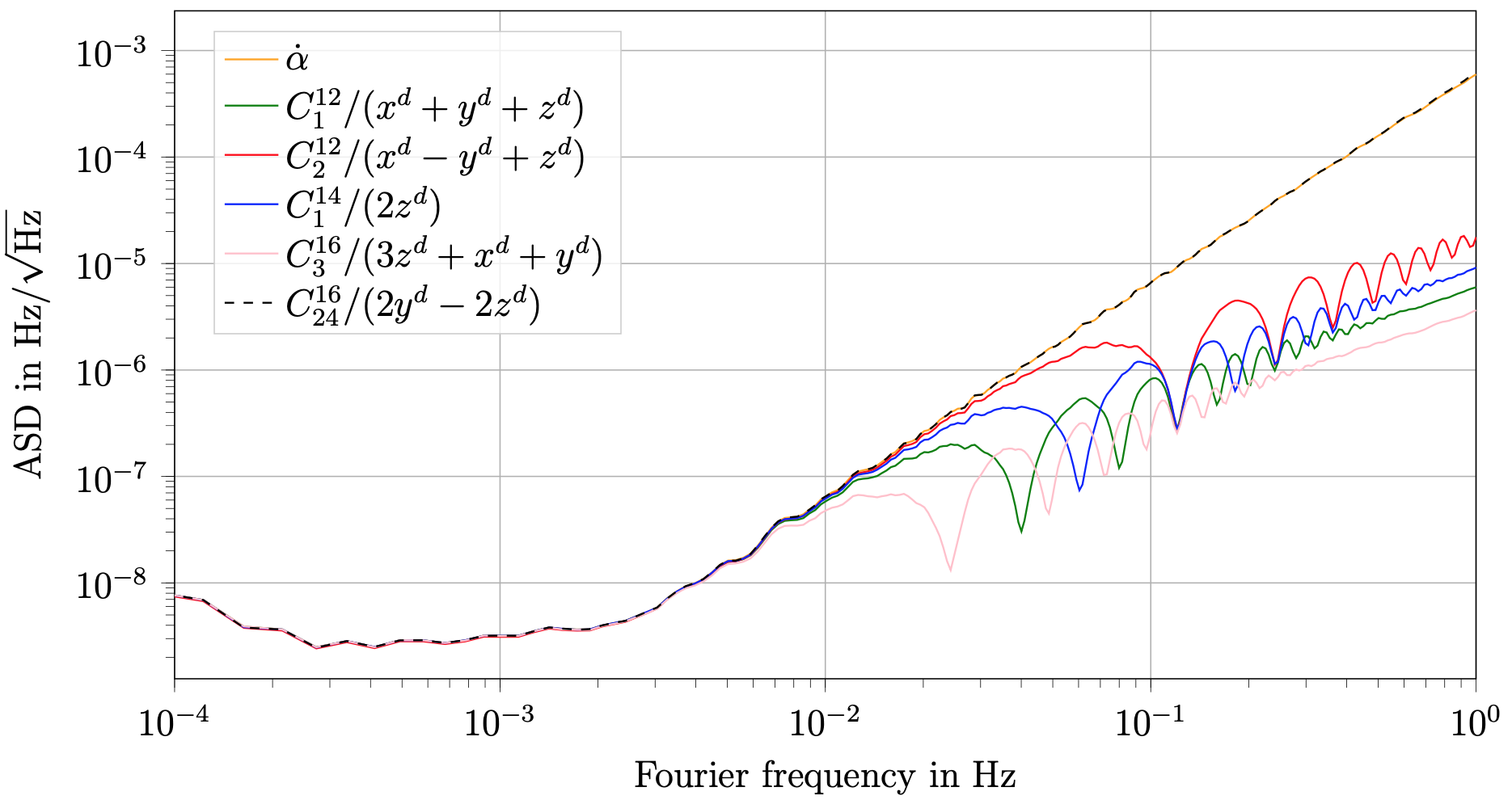}
	\end{center}
\end{figure}

To conclude this section, we compare in \cref{fig:alpha-demo} several possible second generation versions of the generator $\alpha$, which differ by the differential time delays applied to them. As discussed above, these correspond to frequency domain transfer functions of the form $2 \abs{\sin(\pi f \delta)}$, with $\delta$ as the differences of the applied delays.
At low frequencies, we can expand these transfer functions to first order in $\delta$, to get a simple factor $2 \pi f \delta$, which represents a derivative combined with a rescaling by $\delta$. Thus, it is interesting to compare the different possible second generation versions of $\alpha$ to the derivative of the first generation version.
We plot in \cref{fig:alpha-demo} the combinations $C^{12}_1$, $C^{12}_2$, $C^{14}_1$, $C^{16}_3$ and $C^{16}_{24}$, all rescaled by their respective $\delta$, compared to the two-point derivative of $\alpha$ (labelled $\dot \alpha$). We observe that, as expected, all curves coincide at low frequencies, while the different number of zeros determines the deviations at high frequencies. In particular, we observe that $C^{16}_{24}$ seems to approximate $\dot \alpha$ very well across the whole frequency band.\\
However, even if \cref{fig:algebraic-demo} shows that the approximation reported in \cref{tab:all-combs-decomposed} holds within 3 orders of magnitude, we are not able to demonstrate that $C^{16}_{24}$ agrees to $\dot \alpha$ to this precision. The reason is that the two-point derivative we used to compute $\dot \alpha$ has a time difference of $1/f_s = \SI{0.25}{\second}$, which is different to the one of $C_{24}^{16}$, which is $2(z^d - y^d)\approx \SI{0.2}{\second}$. We discuss this in detail in \cref{app:Derivative}.

\subsection{\label{Dsecondgeneration}Simulations including laser noise for the fully symmetric Sagnac variables }

The second generation versions of $\zeta$ given in \cref{tab:all-combs-decomposed}, such as $C_3^{12}$, $C_{27}^{16}$ or $C_{28}^{16}$, are different from the second generation $\zeta_1$ proposed in \cite{second-tdi-2}. For reference, $\zeta_1$ can be written in our notation as
\begin{equation}
	\begin{split}
		\zeta_1 = &(\delay{23}\delay{32} - \delay{13}\delay{21}\delay{32})\eta_{12}
		- (\delay{32}\delay{23} - \delay{12}\delay{31}\delay{23})\eta_{13} \\
		&+ (\delay{32}\delay{13} - \delay{12}\delay{31}\delay{13})(\eta_{23}- \eta_{21}) \\
		&+ (\delay{23}\delay{12} - \delay{13}\delay{21}\delay{12})(\eta_{31} - \eta_{32})  \qc
	\end{split}
\end{equation}

\begin{figure}
	\centering 
	\subfloat[Residual laser noise in $\zeta$ variables in units of \si{\hertz\per\sqrt\hertz}.]{%
		\includegraphics[width=0.9\textwidth]{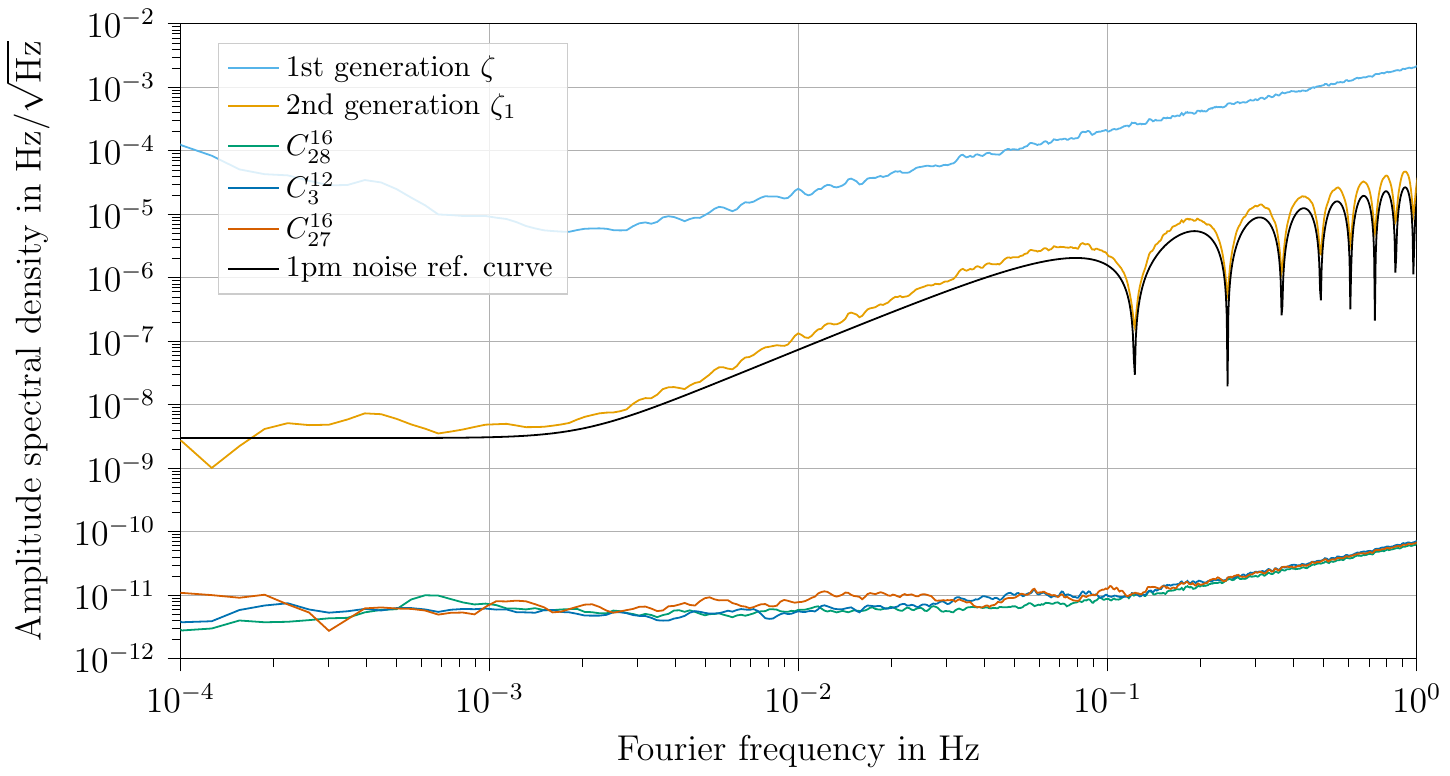}%
		\label{plot:subfig1}%
	}\qquad
	\subfloat[Residual laser noise in $\zeta$ variables in units of \si{\meter\per\sqrt\hertz}.]{%
		\includegraphics[width=0.9\textwidth]{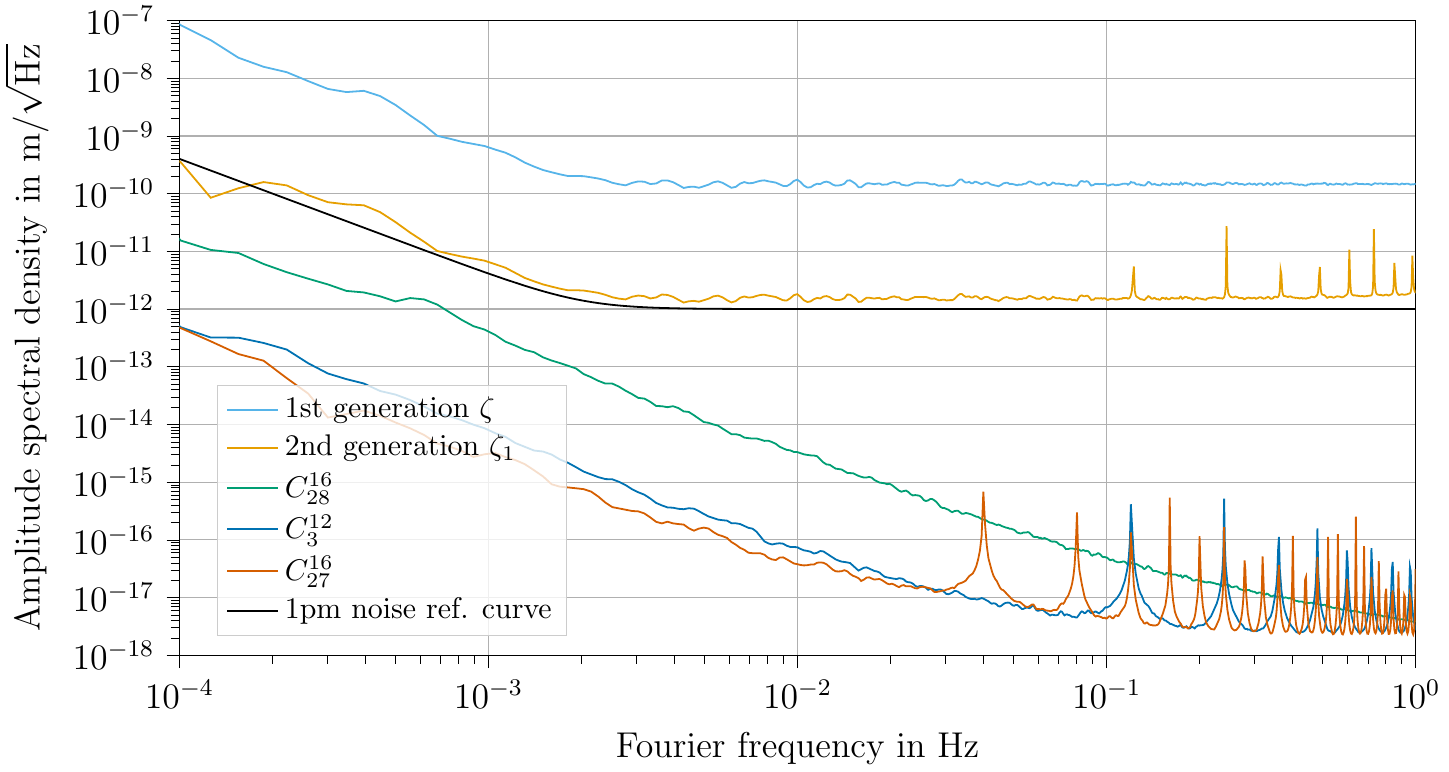}%
		\label{plot:subfig2}%
	}
\caption{Laser noise residuals for 1st and 2nd generation $\zeta$ from the literature compared to $C_{3}^{12}$, $C_{27}^{16}$ and $C_{28}^{16}$ and a \SI{1}{\pico\meter} reference curve. \Cref{plot:subfig1} shows the TDI simulation output in units of \si{\hertz\per\sqrt\hertz}. $C_{3}^{12}$, $C_{27}^{16}$ and $C_{28}^{16}$ perform significantly better than $\zeta_1$, for which the residual laser noise level is above the \SI{1}{\pico\meter} curve, computed accounting for the TDI transfer function of $\zeta_1$. $C_{3}^{12}$, $C_{27}^{16}$ and $C_{28}^{16}$ suppress laser noise below the numerical limit of our simulation for frequencies below \SI{0.1}{\hertz}. \Cref{plot:subfig2} shows the same data, but each variable has been rescaled by their respective transfer function to uncorrelated readout noise in each channel and converted to units of \si{\meter\per\sqrt\hertz}. This allows simultaneous comparison of all variables to the \SI{1}{\pico\meter} curve. \label{laserred}}
\end{figure}

\Cref{laserred} compares the residual laser noise for the 1st and 2nd generation variables $\zeta$ and $\zeta_1$ from the literature with the ones of $C_3^{12}$, $C_{27}^{16}$ and $C_{28}^{16}$. These are given in units of \si{\hertz\per\sqrt\hertz} in \cref{plot:subfig1}, and rescaled to units of \si{\meter\per\sqrt\hertz} in \cref{plot:subfig2}. For reference, we also plot a  typical \SI{1}{\pico\meter\per\sqrt\hertz} reference noise curve\footnote{The overall noise allocation given for the interferometric readout given in \cite{Audley:2017drz} is \SI{10}{\pico\meter\per\sqrt\hertz}. It is common to compare individual noise sources to a more conservative requirement, such that no single noise source uses up the whole allocated noise level.}.
 We can see that the three combinations $C_{3}^{12}$, $C_{27}^{16}$ and $C_{28}^{16}$ reduce laser noise far below the level of the previously known variables (and the requirements), as also demonstrated by theoretical calculation in \cite{Muratore_2020}. As visible in \cref{plot:subfig1}, the residual noise reaches the noise floor of our numerical simulations at a level around \SI{5e-12}{\per\hertz\tothe{0.5}}. Although \cref{plot:subfig1} shows that the variables  $C_{3}^{12}$, $C_{27}^{16}$ and $C_{28}^{16}$ all reach the same level of residual laser frequency noise, they have different transfer functions for secondary noises, such as uncorrelated readout noise in each interferometer. Therefore, the same residual laser noise level should be compared to different residual readout noise levels. We account for this by normalising out the response of each TDI variables readout noise transfer function in \cref{plot:subfig2}, where we also convert the plot to units of \si{\meter\per\sqrt\hertz}, to allow direct comparison of each variables residual laser noise to the standard \SI{1}{\pico\meter} noise reference curve.
 
We observe that the laser noise in TDI $C_{3}^{12}$ and $C_{27}^{16}$ is less significant compared to TDI $C_{28}^{16}$, specifically at low frequencies. The reason is the strong low-frequency suppression of the factor $y^2 - z^2$ present in $C_{28}^{16}$ as given in \cref{tab:all-combs-decomposed}, which applies to readout noise, but not the laser noise. We remark that the residual laser frequency noise level is still significantly below the \SI{1}{\pico\meter} curve across the whole frequency band for all three new variables, even if this transfer function is taken into account.

\clearpage
\section{\label{sec:conclusion} Conclusion}

We have shown explicitly how all 34 second generation core TDI combinations up to 16 links presented in the literature can be expressed in terms of the four first-generation variables $\alpha, \beta, \gamma$ and $\zeta$. We also provide the symmetry rules extending these results to all 210 distinct combinations up to 16 links.

We have verified numerically that these expressions are valid to within 3-5 orders of magnitude, such that a set of four second generation versions of $\alpha, \beta, \gamma$ and $\zeta$ should be sufficient for purposes such as instrumental noise characterisation. The three Sagnac variables $\alpha, \beta$ and $\gamma$ are already enough to construct the quasi-orthogonal channels $A$, $E$ and $T$ often used in LISA data analysis. The variable $\zeta$ can be shown also to be quasi-orthogonal to $A$ and $E$, while having a simpler but otherwise equivalent response to $T$, such that we recommend to use $\zeta$ instead of $T$.  In particular, both $\zeta$ and the $T$ channel are known to be insensitive to GWs at low frequencies, making them useful as noise monitors for the LISA mission, as explored in \cite{muratore2021time}.

One possible set of 2nd generation variables are the combination $C^{12}_{1}$ and its cyclic permutations, combined with one of the new versions of the $\zeta$ variable, such as $C^{16}_{27}$. While $C^{12}_{1}$ and their two cyclic permutation are the standard second generation Sagnac variables known from the literature, $C^{16}_{27}$ (as well as the other variants of $\zeta$ in \cref{tab:all-combs-decomposed}) is new, and significantly more capable of suppressing laser noise then the second generation $\zeta_1$ proposed before \cite{second-tdi-2}.

However, as already discussed in \cite{Vallisneri:2005, Muratore_2020,muratore2021time}, other  combinations might have practical advantages: some, such as the Relay-like variables $C_6^{16}$,$C_7^{16}$,$C_8^{16}$, or the Monitor-like variables $C_{21}^{16}$ and $C_{22}^{16}$, use just 4 out of the full 6 laser links, such that they are unaffected in case of a loss of one or two inter-satellite links. Others, such as the Michelson-like combinations $C_1^{16}$, $C_{4}^{16}$ and $C_{5}^{16}$, use just 4 inter-satellite links and in addition just 2 constellation arms, thus they remain available in case of a complete failure of one of the LISA arms.\\ Moreover, as we illustrated in \cref{fig:alpha-demo}, the typical singularities present in the transfer function of all second generation TDI variables appear at different frequencies for the different combinations. This might allow simplified data analysis pipelines for some combinations which lack zeros inside the main LISA band. In particular, one such set of combinations we found are $C^{12}_{2}$, its cyclic permutations, and $C^{12}_{3}$.

Special mention should be given to the combinations $C^{16}_{4}$, $C^{16}_{24}$, and $C^{16}_{28}$, which are exactly vanishing under the assumption of equal arms, and completely lack additional zeros below \SI{1}{\hertz} when realistic orbits are considered. Despite this interesting property, they show to be unstable under changes in the orbital dynamics, such that we are hesitant to recommend their use for most data analysis purposes.

Last but not least, combinations with multiple measurements require shorter segments of data to compute a single data point of the TDI combination. For instance, each of the two beams of $C_1^{12}$ require summation of the light travel time for 6 consecutive links, or about $6\times \SI{8.33}{\second} \approx \SI{50}{\second}$. The beams of $C_2^{12}$, on the other hand, use only up to 4 links at a time, corresponding to just $4\times \SI{8.33}{\second} \approx \SI{33}{\second}$. \\
We remark that the new second generation combinations presented in \cite{Vallisneri:2005,Muratore_2020,muratore2021time} and studied here require a slightly more complicated (but well understood) implementation of the TDI algorithm, since they utilize not only delays, but also time-advancements. In addition, the 'classic' 2nd generation TDI variables, such as $C_1^{12}$ and its cyclic permutations, allow a conscise and numerically efficient factorization of the different time-shifts, which might not always be possible to the same extend for the new variables. However, we do not expect these points to ultimately impact the data analysis for LISA.\\
Note that we did not investigate the case of 1.5th generation TDI using 6 generators, as we notice that the 1st generation was enough to accurately reproduce the response of all core TDI variables with respect to instrumental noise. Still, it might be valuable as a follow up study to see if we can extract additional information in this case.

To conclude, based on these arguments, and assuming all interferometer arms remain fully operational, we would recommend to investigate the use of $C^{12}_{2}$, its cyclic permutations, and $C^{12}_{3}$ as basis for the LISA data analysis. $C^{12}_{2}$ and its cyclic permutations can be used to construct versions of the variables $A$ and $E$, which together with $C^{12}_{3}$ as second generation version of $\zeta$ form a quasi-orthogonal set.

\section{Acknowlegement}
We thank the LISA Trento group for the fruitful discussion, in particular S. Vitale and D. Vetrugno. We also want to thank M. Staab from the AEI in Hannover for the useful comments on improving this manuscript. In addition, we thank the SYRTE theory and metrology group, S. Babak and an anomynous referee for useful remarks in improving an initial version of this manuscript. M.M thanks the Agenzia Spaziale Italiana and the Laboratorio Nazionale di Fisica Nucleare for supporting this work. O.H. gratefully acknowledges support by the Deutsches Zentrum f\"ur Luft- und Raumfahrt (DLR, German Space Agency) with funding from the Federal Ministry for Economic Affairs and Energy based on a resolution of the German Bundestag (Project Ref.~No.~50OQ1601 and 50OQ1801). O.H. also acknowledges funding from Centre National des Etudes Spatiales.

\clearpage
\appendix

\section{Time shift operators}
\label{sec:notations}

This paper makes use of time shift operators. They act on time dependent functions by evaluating them at another time. 
We thus define the following notations related to time-shift operators and TDI combinations:

\begin{itemize}
	\item{Delay operator: $D_{ij}\phi_{j}(\tau) = \phi_{j}(\tau - d_{ij}(\tau))$.
	
	Given a time of reception $\tau$ of a beam on spacecraft $i$, evaluates the phase $\phi_{j}$ of that beam at the time of emission at spacecraft $j$, which we write as $\tau-d_{ij}(\tau)$. Note that depending on  what frame $\phi_{j}(\tau)$ is defined in, the computation of $d_{ij}$ can include a change in reference frames, and clock offsets.
	}
	\item{Advancement operator: $A_{ij} \phi_{j}(\tau) = \phi_{j}(\tau + a_{ij}(\tau))$.
	
	Given a time of emission $\tau$ of a beam from spacecraft $j$, evalutes the phase $\phi_{j}$ of that beam at the time of reception on spacecraft $i$, which we write as $\tau+a_{ij}(\tau)$. This is the inverse operation to that of the delay operator, such that we have the identity $A_{ij}D_{ji} \phi_{i}(t) = D_{ij}A_{ji} \phi_{i}(t) =  \phi_{i}(t)$.
}

\item{Multiple Delay operators: $D_{ij} D_{jk}\phi_{k}(\tau) = \phi_{k}(\tau-d_{ij}(\tau) - d_{jk}(\tau-d_{ij}(\tau)).$}

\item{Multiple Delay and Advancement operators: $A_{n i}  D_{ij} D_{j k}\phi_{k}(\tau) =\phi_{k}(\tau+ a_{ni}(\tau)-d_{ij}(\tau+ a_{ni}(\tau))-d_{jk}(\tau+ a_{ni}(\tau)-d_{ij}(\tau+ a_{ni}(\tau)))) .$}
	\end{itemize} 
 Only the delays $d_{ij}(\tau)$ are directly accessible from the LISA measurements. The advancements $a_{ij}(\tau)$ can be computed from them by iteratively solving
 \begin{equation}
 	a_{ij}(\tau) = d_{ji}(\tau + a_{ij}(\tau)),
 \end{equation}
 which directly follows from $A_{ij}D_{ji} \phi_{i}(t) =  \phi_{i}(t)$.
\section{Demonstration of validity of decomposition in the presence of laser noise}
\label{decompositionofTDI}
As a proof of concept that the decompositions presented in \cref{tab:all-combs-decomposed} are still applicable when we include laser noise in the simulation, we want to linearly combine multiple second generation variables to construct a version of the variable $C_1^{16}$.

The previously discussed set of variables without zeros in the LISA band seem to be good candidates to use in this construction. However, we remark that the two cyclic permutations of $C_{24}^{16}$ contain the time differences $2(x^d - z^d)$ and $2(y^d - z^d)$, respectively, such that including $C_{24}^{16}$ itself, we get finite difference approximations of $\dot \alpha, \dot \beta$ and $\dot \gamma$ with unequal time differences. The same argument holds for $C_{28}^{16}$, which approximates $\dot \zeta$. This limits our ability in using linear combinations of these variables to build other variables in the table, as discussed in the \cref{app:Derivative} below and illustrated in \cref{fig:res-X}.\\
On the other hand, as visible in \cref{tab:all-combs-decomposed}, the variables $C^{12}_{1}$ (plus its cyclic permutations $\hat{C}_1^{12}$ and $\hat{\hat{C}}_1^{12}$) and $C^{16}_{27}$ all have the same time shift factor $(1 - xyz)$ applied to them. This allows us to linearly combine them without introducing additional errors.

Let us define a short-hand notation for the second generation representatives of the variables $\alpha$, $\beta$, $\gamma$ and $\zeta$ chosen for this example:
\begin{subequations}
	\begin{align}
		\tilde \alpha & \equiv C_1^{12} \label{eq:string2-alpha}\\
		\tilde \beta &\equiv \hat{C}_1^{12} \label{eq:string2-beta}\\
		\tilde \gamma &\equiv\hat{\hat{C}}_1^{12} \label{eq:string2-gamma}\\
		\tilde \zeta &\equiv x C_{27}^{16}.\label{eq:string2-zeta}
	\end{align}
\end{subequations}
Applying the factor $(1 - xyz)$ to the expression given for $C_1^{16}$ in \cref{tab:all-combs-decomposed}, we see that
\begin{subequations}
	\begin{align}
		(1 - x y z) C_1^{16} & \approx \left(1-y^2 z^2\right) (1 - x y z)  \qty(\alpha -z  \beta -y \gamma +y z \zeta)  \\ & \approx  \left(1-y^2 z^2\right)  \qty(\tilde \alpha -z \tilde  \beta -y \tilde \gamma +y z \tilde \zeta). \label{eq:c1-16-approx}
	\end{align}
\end{subequations}
To verify that \cref{eq:c1-16-approx} is accurate, we compute the exact version of $C_1^{16}$ given in \cref{tab:core-combinations-in-eta} numerically, and apply the additional factor $(1 - xyz)$ to the resulting laser-noise free variable. We compare it to the approximated version on the right hand side of \cref{eq:c1-16-approx}, where we compute $\tilde\alpha$, $\tilde\beta$, $\tilde\gamma$ and $\tilde\zeta$ as given in \cref{eq:string2-alpha,eq:string2-beta,eq:string2-gamma,eq:string2-zeta}.

\begin{figure}
	\caption{Secondary noise levels in  $(1 - x y z)C^{16}_1$ compared to
	 the residual between the approximation given in \cref{eq:c1-16-approx} and the exact expression given in \cref{tab:core-combinations-in-eta}. Laser noise is included in the simulation, but fully suppressed by TDI. In addition, we give an analytical estimate of the expected level of the secondary noise levels.
			\label{fig:algebraic-demo2}}
	\begin{center}
	\includegraphics[width=\textwidth]{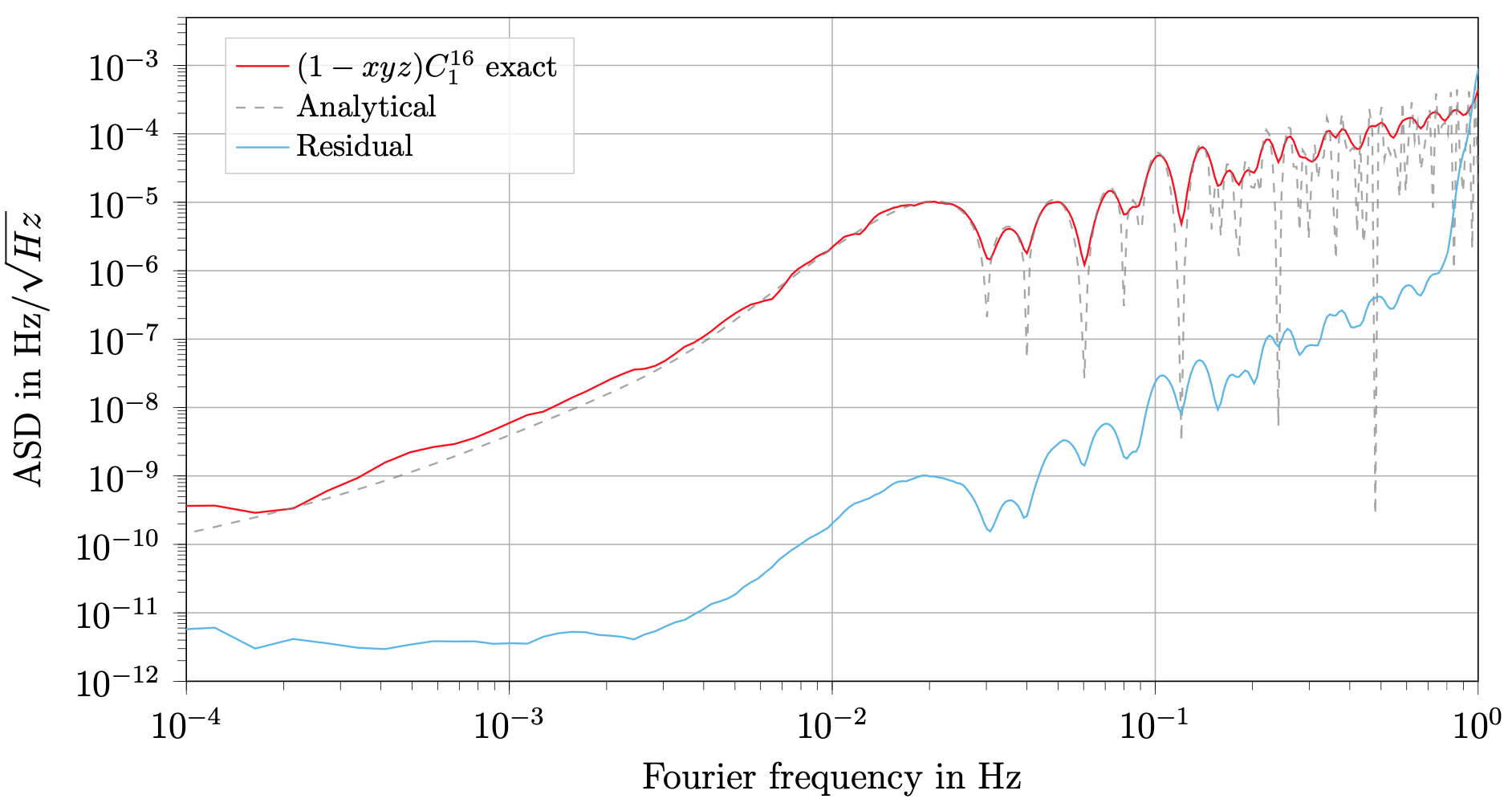}
	\end{center}
\end{figure}

 \Cref{fig:algebraic-demo2} shows that the noise level of the simulated data (in orange) is well explained by an analytical model describing the secondary noises (in dotted grey). For clarity the approximated solution of $(1 - xyz)C_1^{16}$ is omitted as we cannot appreciate the difference with respect to the exact solution on this scale. We show instead that the residual noise between the left- and right-hand side of \cref{eq:c1-16-approx} is several orders of magnitude below the secondary noises. This same principle could easily be applied to any of the variables given in \cref{tab:all-combs-decomposed}. \\

\section{Numerical approximations of $\dot \alpha$ }
\label{app:Derivative}
\Cref{fig:alpha-demo} showed that $C_{24}^{16}$ seems to be a good approximation for the derivative of the first generation variable $\alpha$ accross the whole LISA frequency band. To quantify this, we want to study the residual between $C_{24}^{16}$ and the two point finite difference of $\alpha$, computed directly from the first-generation variable, which we denote by $\dot \alpha_{2p}$.

\begin{figure}
	\caption{Comparison between the estimation of a two-point finite difference derivative of $\alpha$ and ${C}^{16}_{24}$. 	\label{fig:res-alpha}}
	\begin{center}
	\includegraphics[width=\textwidth]{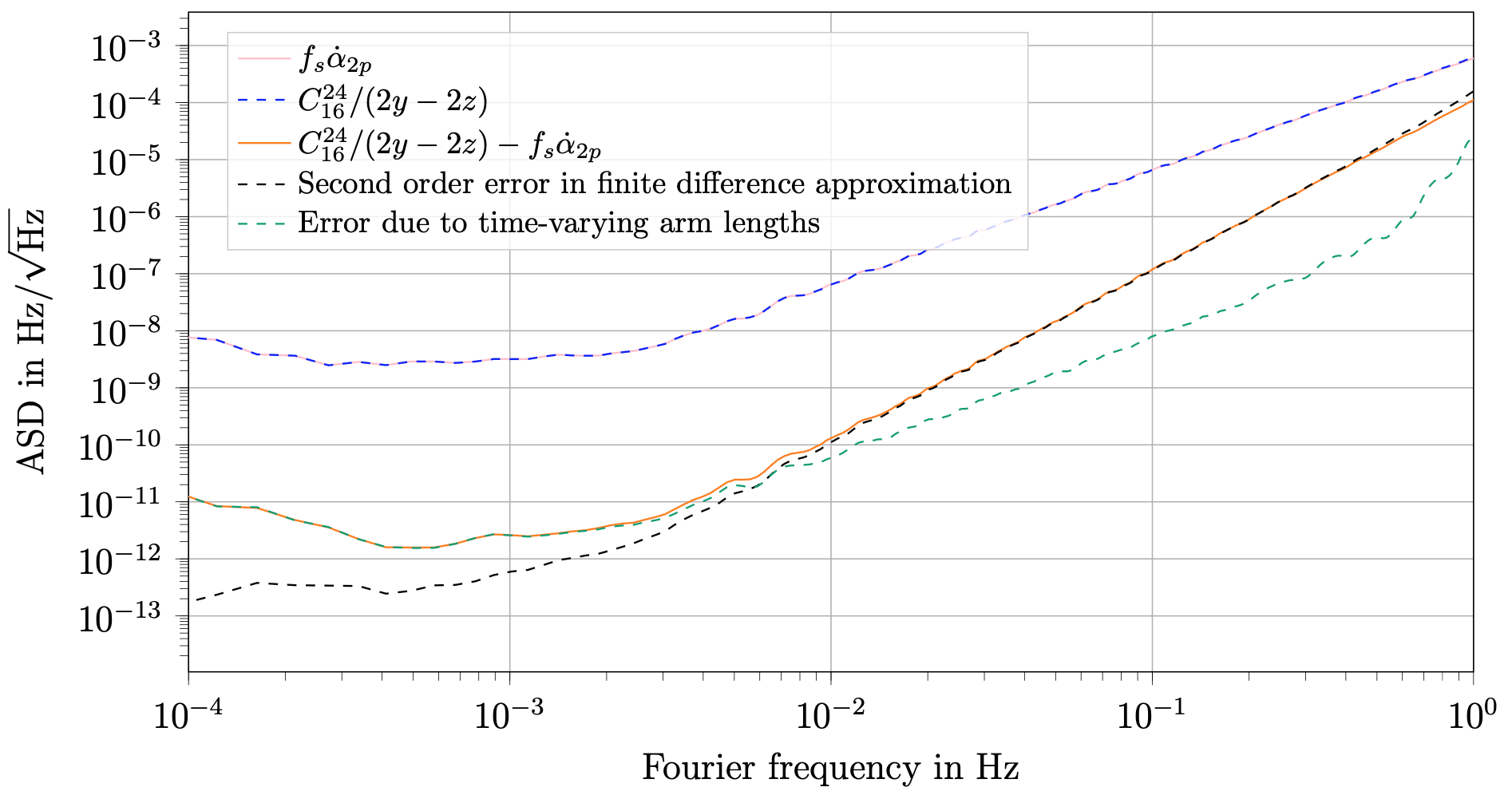}
	\end{center}
\end{figure}

\Cref{fig:res-alpha} shows the comparison between $\dot \alpha_{2p}$ and the expression of ${C}^{16}_{24}$ reported in \cref{tab:all-combs-decomposed}. The plot shows also the residuals between the two numerical computations and a model which explains their values. 
While the computations agree within three orders of magnitudes at low frequencies, the error increases
towards higher frequencies where the residuals reach about one order of magnitude below the actual value.

This behaviour of the residuals can be explained by two separate effects. For the high frequencies range, we have to take into account the inequality between the time differences, $\delta$, we consider to approximate the derivative of $\alpha$ in $\dot \alpha_{2p}$ and ${C}^{16}_{24}$, respectively. \\

In both cases, we have a finite difference of the form $\frac{\alpha(t) - \alpha(t - \delta)}{\delta}$ which we can expand to first order in $\delta$ to get:
\begin{equation}
	\frac{\alpha(t) - \alpha(t - \delta)}{\delta} \approx \dot \alpha(t) -  	\frac{\delta}{2} \ddot \alpha(t),
\end{equation}
where $\delta = 2(y^d-z^d)$ for ${C}^{16}_{24}$  and $\delta = 1/f_s$ for $\dot \alpha_{2p}$, while $f_s = \SI{4}{\hertz}$ is the sampling frequency. 

Thus the difference between the two approximated derivatives will be given by:
\begin{equation}
\frac{{C}^{16}_{24}}{2(y^d-z^d)} -  f_s\dot {\alpha}_{2p}(t) \approx \left( \frac{1}{2f_s} - y^d +z^d \right)  \ddot{\alpha}(t).
\end{equation}
In the frequency domain, the additional derivative corresponds to a factor $2\pi f$, which explains the increase of the residuals at high Fourier frequencies.\\ 

Regarding the low frequencies, the residuals that we see are explained by the error that we make in estimating the ${C}^{16}_{24}$ variable out of the TDI $\alpha$ first generation as visible from \cref{fig:algebraic-demo}. 
We estimate it by rescaling the PSD of $\dot\alpha$ by the ratio between the residual estimated in \cref{fig:algebraic-demo} and the actual PSD of ${C}^{16}_{24}$.

The same reasoning holds for $\dot \beta$, $\dot \gamma$ using the cyclic permutations of ${C}^{16}_{24}$, and $\dot \zeta$ using ${C}^{16}_{28}$.

\begin{figure}
	\caption{Comparison between the two-point derivative of the combination ${C}^{16}_1$ in \cref{tab:core-combinations-in-eta} and the version built out of rescaled versions of $C_{24}^{16}$, its cyclic permutations, and $C_{28}^{16}$, which represent $\dot \alpha$, $\dot \gamma$, $\dot \beta$, and $\dot \zeta$, respectively. \label{fig:res-X}}
	\begin{center}
	\includegraphics[width=\textwidth]{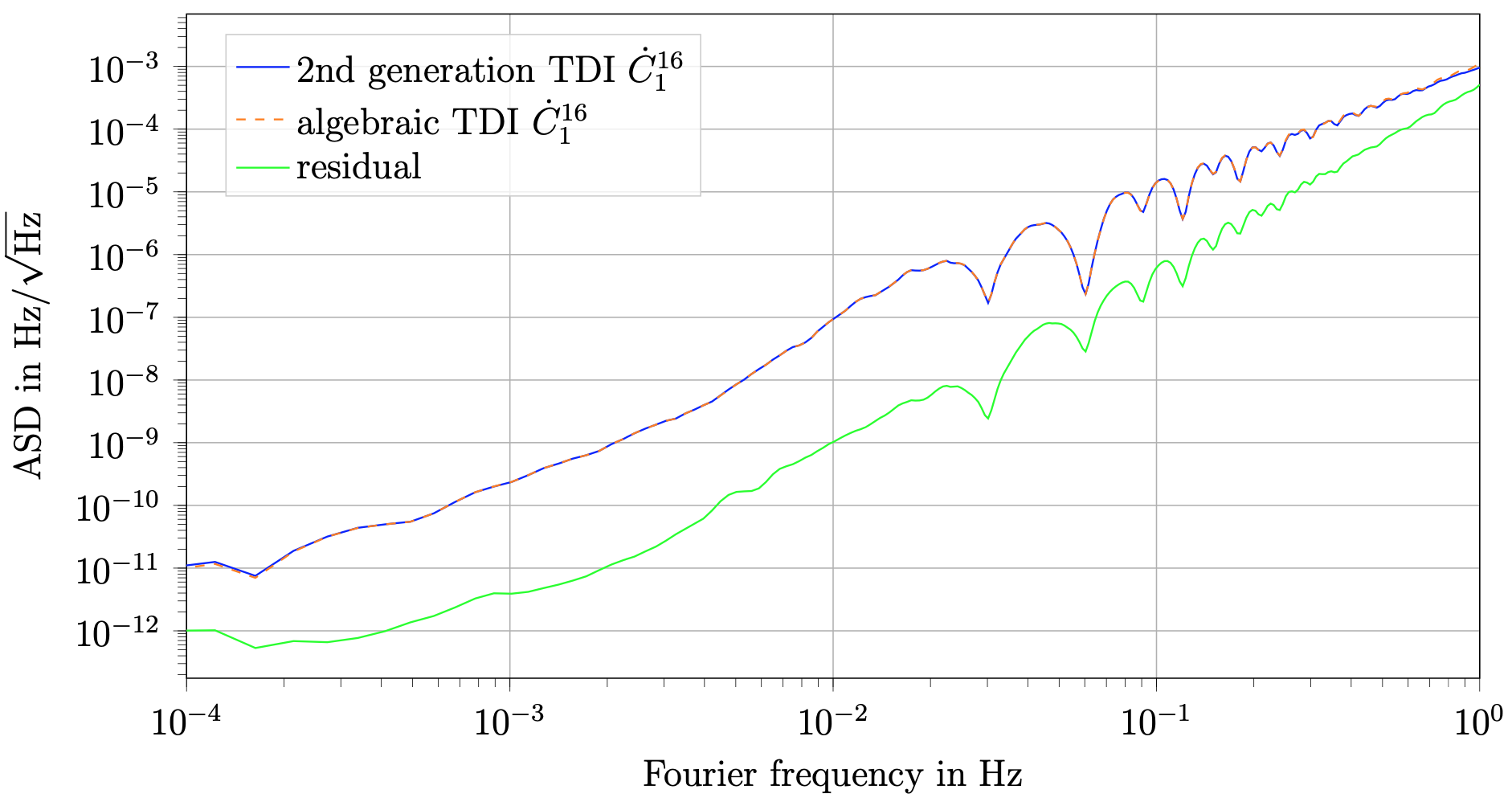}
%		\makebox[\textwidth][c]{\input{./tikz/res-X.tex}}
	\end{center}
\end{figure}

We can then use these variables in place of $\alpha$, $\beta$, $\gamma$ and $\zeta$ to build an approximate version of the derivative of ${C}^{16}_{1}$, using the expression given in \cref{tab:all-combs-decomposed}. \Cref{fig:res-X} shows the comparison between $\dot{C}^{16}_{1}$ computed using a two-point derivative and the version computed using ${C}^{16}_{24}$, ${\hat C}^{16}_{24}$, $\hat{\hat{C}}_{24}^{16}$ and ${C}^{16}_{28}$, as well as the relative difference between the two calculations. We can see how the residuals are two orders of magnitude lower than the variable we are trying to compute and that they increase at higher frequencies. This is in accordance to the two complementary models described above.

\section{Symmetries}
\label{app:symmetries}

In this paper, we study only the core combinations from which other variables can be constructed. We summarize here how to apply these symmetries to the decompositions presented in \cref{tab:all-combs-decomposed}.

\subsection{Cyclic permutatation}
A cyclic permutation maps the spacecraft indices from $1\mapsto 2\mapsto 3 \mapsto 1$. This corresponds to the following mappings in \cref{tab:all-combs-decomposed}:
\begin{itemize}
	\item Map $\alpha\mapsto \beta\mapsto \gamma \mapsto \alpha$,
	\item Map $x\mapsto y\mapsto z \mapsto x$,
	\item Leave the fully symmetric $\zeta$ unchanged.
\end{itemize}

\subsection{Mirror symmetry}
A mirror symmetry exchanges the role of two spacecraft, for example  $2\leftrightarrow 3$. This corresponds to the following mappings in \cref{tab:all-combs-decomposed}:
\begin{itemize}
	\item Exchange $\beta\leftrightarrow \gamma$,
	\item Exchange $y\leftrightarrow z$,
	\item Flip the sign of all combinations.
\end{itemize}
Similar rules apply for the reflections involving spacecraft $1\leftrightarrow 3$ and $1\leftrightarrow 2$, which leave either $\beta$ and $y$ or $\gamma$ and $z$ unchanged, respectively.

\subsection{Time reversal symmetry}
The action of a time reversal of the combination is less obvious than the previous two symmetries. Note that a time reversal is equivalent to one of the other two symmetries for most variables. For the ones were this is not the case ($C_1^{14}, C_{11}^{16},C_{15}^{16}, C_{17}^{16}, C_{21}^{16}, C_{22}^{16}, C_{24}^{16}$ and $C_{28}^{16}$, cf. \cite{muratore2021time}), we computed the corresponding expression to verify if these variables bring additional information.  It turns out that in the approximations of this paper, a time reversal reduces to one of the other symmetries in most cases, plus an additional overall time shift and sign flip. The exceptions are $C_{21}^{16}$ and $C_{22}^{16}$, whose time reversed versions $C_{21}^{\text{tr},16}$ and $C_{22}^{\text{tr},16}$ have the decompositions
\begin{align}
	x^2 y z C_{21}^{\text{tr},16} &= (y - xz)(z\gamma - \zeta), \\
	x z^2 C_{22}^{\text{tr},16} &= (z^2 - 1)(z\gamma - \zeta). 
\end{align}
As mentioned in \cref{ssec:theory-discussion}, $C_{21}^{16}$ and $C_{22}^{16}$ have been identified already in \cite{Vallisneri:2005} as 'Monitor'-type variables, in which one spacecraft only acts as receivers. Applying the time reversal transforms them to the 'Beacon'-type variables found in \cite{Vallisneri:2005}, in which the same spacecraft only acts as emitter, which explains why they have a different representation.

\section{Quantitative example of the noise added at interferometer nulls}
\label{app:laser-noise-at-zeros}
As argued in \cref{Dfirstgeneration}, the additional zeros present in second generation TDI variables, which theoretically do not affect the SNR, will be filled up by other noise sources which have a different transfer functions, such that we expect a slight degradation of SNR close to these frequencies in practice.

In this paper, we focus on laser noise and the (so-called) secondary noises that are readout and test mass acceleration noise. Test-mass noise is sub-dominant at the higher frequencies where the additional zeros appear. Thus, to quantify the bandwidth we loose for the presence of the zeros, let us consider the analytical models of the readout noise contributions,  expressed as an amplitude spectral density. We do this for the example of the two combinations $C^{12}_3$ and $C^{16}_{27}$ under the assumption of equal arm lengths, which we write following \cref{tab:all-combs-decomposed} as
\begin{equation}
	C^{12}_3 \approx (1 - D)\zeta\qcomma C^{16}_{27} \approx (1 - D^3)\zeta.
\end{equation}
Assuming readout noise to be uncorrelated and of equal magnitude in each measurement $\eta_{ij}$, it immediately follows from \cref{eq:operator-zeta} that the readout noise level in $\zeta$ is just $\sqrt{6 S_{oms}}$. For simplicity, we assume that our readout noise level is equal to the overall noise level of the optical metrology system (OMS) in a single link as given in \cite{LISAPM}. Expressed in units of frequency, we have $S_{\text{oms}} = \qty(\frac{\SI{12}{\pico\meter}}{\si{\sqrt\hertz}}\frac{2\pi f}{\SI{1064}{\nano\meter}})^2$.

The factors $(1 - D)$ and $(1 - D^3)$ correspond to an additional sinusoidal transfer function applied to the response of $\zeta$, such that we get
\begin{equation}
C^{12, OMS}_3 \approx \qty|2 \sin(T \pi f)| \sqrt{6 S_{oms}},\label{Eq:omsmodel1}
\end{equation}
and
\begin{equation}
C^{16, OMS}_{27} \approx \qty|2 \sin(3T \pi f)| \sqrt{6 S_{oms}}.\label{Eq:omsmodel2}
\end{equation}
Since we are interested in estimating the level of the residual noise around the nulls for these channels, we can perform a first order series expansion of \cref{Eq:omsmodel1,Eq:omsmodel1} around the argument of the sines around their first zero. We get:
\begin{equation}
	\qty|2\sqrt{6} \sin(T \pi f)| \sqrt{S_{oms}} \approx 2\sqrt{6}\pi\qty|T f - 1|\sqrt{S_{oms}},\label{eq:A1}
\end{equation}
and
\begin{equation}
	\qty|2\sqrt{6} \sin(3T \pi f)| \sqrt{S_{oms}} \approx 2\sqrt{6}\pi\qty|3 T f - 1|\sqrt{S_{oms}}.\label{eq:A2}
\end{equation}
 The residual laser frequency noise in a TDI variable, on the other hand, is proportional to the arm length mismatch,
  \begin{equation}
  2 \pi f \Delta T \sqrt{S_\nu}, \label{eq:A3}
\end{equation}
with $\sqrt{S_\nu} = \SI{28.8}{\hertz\per\sqrt\hertz}$ as the level of the laser frequency noise and $\Delta T \approx \SI{1}{\pico\second}$ \cite{Muratore_2020}. We can equate the right-hand side of both \cref{eq:A1,eq:A2} with \cref{eq:A3}, and solve both equations for $f>0$. This gives the upper and lower bounds around the first zero for which the laser noise is dominant, which allows us to compute the bandwidth $\Delta f$ we lose. \\ 
The first null for $C^{12}_3$ is at $f\approx \SI{0.12}{\hertz}$, and laser noise is dominant in a bandwidth of $\Delta f \approx 4.0031\cdot 10^{-8}$. For $C^{16}_{27}$, the first zero is at $f\approx \SI{0.04}{\hertz}$, and the corresponding bandwidth we lose due to laser noise is just $\Delta f \approx 1.33437 \cdot10^{-8}$. \\ Note that since $S_{oms}$ and the residual laser noise given in \cref{eq:A3} have the same slope, these values are identical for all additional zeros. This means that even though $C^{16}_{27}$ has three times as many zeros as $C^{12}_{3}$, the overall bandwidth that we loose to laser frequency noise is the same for both combinations, since the value of $\Delta f$ for $C^{16}_{27}$ is three times as big as that of $C^{12}_{3}$. In addition, the computed values for $\Delta f$ only make out a tiny fraction of the overall LISA observation band, such that this effect is probably negligable for most applications.

We remark that other fundamental noise sources might lead to higher residuals then the laser frequency noise considered here. \\

One set of such noise sources could be numerical noise and artifacts entering during the TDI processing and the subsequent astrophysical data analysis. For example, even simple data analysis methods such as the spectral estimations used in this paper can provide erroneous results far above the level computed for the laser noise above. This is visible in \cref{fig:algebraic-demo2}, where our analytical model clearly diverges from the estimated spectrum in a significant bandwidth around the zeros, at a much higher level then the limits of our approximations given by the blue line would suggest. The precise level of such errors depends on the implementation and parameters used in the final data analysis algorithms for LISA, which is why we do not attempt to quantify such effects here.
In addition, at least in principle, any on-ground processing should be implementable in such a way that numerical artifacts are not limiting, e.g., by using specialized tools or libraries with increased numerical precision.

\clearpage
\bibliography{biblio.bib}

\end{document}